\documentclass[sigconf]{acmart}
\usepackage{booktabs} %
\usepackage{wrapfig}

\usepackage[ruled]{algorithm2e} %

\SetAlFnt{\small}
\SetAlCapFnt{\small}
\SetAlCapNameFnt{\small}
\SetAlCapHSkip{0pt}

\copyrightyear{2023}
\acmYear{2023}
\setcopyright{acmlicensed}\acmConference[SIGGRAPH '23 Conference Proceedings]{Special Interest Group on Computer Graphics and Interactive Techniques Conference Conference Proceedings}{August 6--10, 2023}{Los Angeles, CA, USA}
\acmBooktitle{Special Interest Group on Computer Graphics and Interactive Techniques Conference Conference Proceedings (SIGGRAPH '23 Conference Proceedings), August 6--10, 2023, Los Angeles, CA, USA}
\acmPrice{15.00}
\acmDOI{10.1145/3588432.3591498}
\acmISBN{979-8-4007-0159-7/23/08}

\begin{document}
\title[Mesh Density Adaptation for Template-based Shape Reconstruction]
    {Mesh Density Adaptation for \\ Template-based Shape Reconstruction}

\author{Yucheol Jung}
\authornote{Equal contribution}
\email{ycjung@postech.ac.kr}
\author{Hyomin Kim}
\authornotemark[1]
\email{min00001@postech.ac.kr}
\affiliation{%
 \institution{POSTECH}
 \city{Pohang}
 \country{South Korea}}

\author{Gyeongha Hwang}
\email{ghhwang@yu.ac.kr}
\affiliation{%
 \institution{Yeungnam University}
 \city{Gyeongsan}
 \country{South Korea}}

\author{Seung-Hwan Baek}
\email{shwbaek@postech.ac.kr}
\author{Seungyong Lee}
\email{leesy@postech.ac.kr}
\affiliation{%
 \institution{POSTECH}
 \city{Pohang}
 \country{South Korea}}
 
\makeatletter
\let\@authorsaddresses\@empty
\makeatother

\begin{CCSXML}
<ccs2012>
   <concept>
       <concept_id>10010147.10010178.10010224.10010245.10010254</concept_id>
       <concept_desc>Computing methodologies~Reconstruction</concept_desc>
       <concept_significance>500</concept_significance>
       </concept>
   <concept>
       <concept_id>10010147.10010371.10010396.10010397</concept_id>
       <concept_desc>Computing methodologies~Mesh models</concept_desc>
       <concept_significance>300</concept_significance>
       </concept>
 </ccs2012>
\end{CCSXML}

\ccsdesc[500]{Computing methodologies~Reconstruction}
\ccsdesc[300]{Computing methodologies~Mesh models}

\keywords{Inverse rendering, diffusion re-parameterization, non-rigid registration, Laplacian regularization}

\newcommand{\Eq}[1]  {Eq.~(\ref{eq:#1})}
\newcommand{\Eqs}[1] {Eqs.~(\ref{eq:#1})}
\newcommand{\Fig}[1] {Fig.~\ref{fig:#1}}
\newcommand{\Figs}[1]{Figs.~\ref{fig:#1}}
\newcommand{\Tbl}[1]  {Table~\ref{tbl:#1}}
\newcommand{\Tbls}[1] {Tables~\ref{tbl:#1}}
\newcommand{\Sec}[1] {Sec.~\ref{sec:#1}}
\newcommand{\Secs}[1] {Secs.~\ref{sec:#1}}
\newcommand{\Etal}{{\textit{et~al.}}}
\newcommand{\Alg}[1] {Algorithm~\ref{alg:#1}}

\newcommand{\setone}[1] {\left\{ #1 \right\}} %
\newcommand{\settwo}[2] {\left\{ #1 \mid #2 \right\}} %

\newcommand{\todo}[1]{{\textcolor{blue}{TODO: #1}}}
\newcommand{\change}[1]{{\color{red}#1}}
\newcommand{\sean}[1]{{\textcolor{magenta}{sean: #1}}}
\newcommand{\jung}[1]{{\textcolor{magenta}{jung: #1}}}
\newcommand{\kim}[1]{{\textcolor{magenta}{kim: #1}}}
\newcommand{\hwang}[1]{{\textcolor{magenta}{hwang: #1}}}
\newcommand{\baek}[1]{{\textcolor{magenta}{baek: #1}}}

\begin{teaserfigure}
    \centering
    \includegraphics[width=0.86\textwidth]{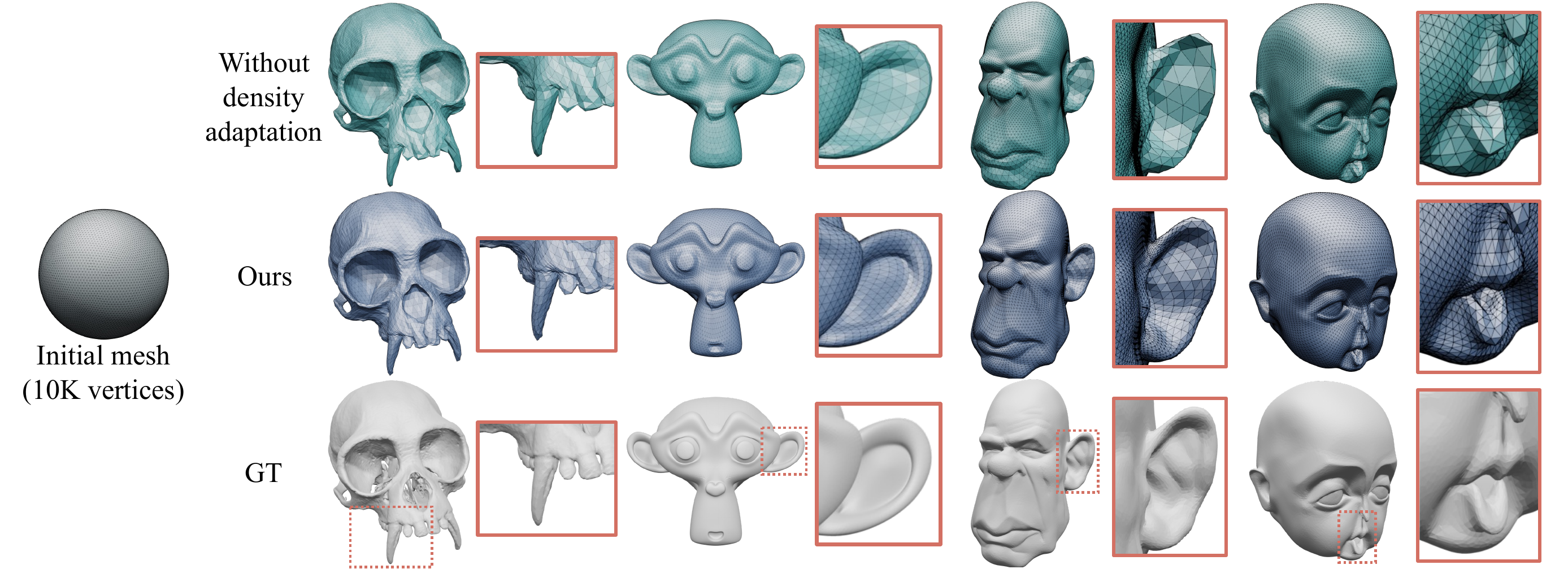}
    \caption{Our mesh density adaptation improves reconstruction quality for complex structures in inverse rendering (first two columns) and non-rigid surface registration (last two columns) 
    while retaining the original template meshing.}
    \label{fig:teaser}
\end{teaserfigure}

\begin{abstract}

In 3D shape reconstruction based on template mesh deformation, a regularization, such as smoothness energy, is employed to guide the reconstruction into a desirable direction. In this paper, we highlight an often overlooked property in the regularization: the vertex density in the mesh. Without careful control on the density, the reconstruction may suffer from under-sampling of vertices near shape details. We propose a novel mesh density adaptation method  to resolve the under-sampling problem. Our mesh density adaptation energy increases the density of vertices near complex structures via deformation to help reconstruction of shape details. We demonstrate the usability and performance of mesh density adaptation with two tasks, inverse rendering and non-rigid surface registration. Our method produces more accurate reconstruction results compared to the cases without mesh density adaptation. Our code is available at \url{https://github.com/ycjungSubhuman/density-adaptation}.

\end{abstract}

\maketitle

\section{Introduction}

3D surface reconstruction is a task where a 3D object or scene is obtained from observations. 
Various approaches have been studied for surface reconstruction based on parametric models \cite{blanz1999morphable}, implicit functions \cite{kazhdan2013screened}, and point set triangulation \cite{cazals2006delaunay}. In this paper, we focus on the template-based approach for surface reconstruction.

In template-based surface reconstruction, a 3D surface is represented as a deformed template mesh. This approach has been widely adopted in non-rigid surface registration frameworks that target 3D reconstructions in specific domains, such as face \cite{li2017learning} and human body \cite{bogo2014faust}. In the frameworks, a template mesh is deformed to fit multiple scans, and the template mesh provides a shared parameter domain for the scans. Then, the dense correspondence between the scans is trivially defined, enabling construction of data-driven parametric models \cite{paysan20093d} and texture transfer \cite{allen2003space} between the models.

Template-based surface reconstruction is also used for reconstructing 3D shapes in inverse rendering problems. In \cite{kato2018neural, nicolet2021large}, a sphere is used as the template mesh that provides effective discretization for a general class of genus-0 shapes. Using the template mesh, surface reconstruction from images reduces to updating 3D positions of template mesh vertices to match the rendering results. The deformed template with updated vertex positions is rendered using a differentiable renderer \cite{laine2020modular} and is optimized via gradient-based optimization.

Template-based surface reconstruction is usually formulated as an iterative optimization of an energy of the form $E_g(\mathbf{p}, \mathbf{o})$, where $\mathbf{p} \in \mathbb{R}^{N \times 3}$ is template vertex position and $\mathbf{o}$ is observation, e.g., images or a given 3D surface. $E_g$ defines the primary goal of the reconstruction, e.g., photometric loss in inverse rendering or Chamfer distance in non-rigid registration. However, as the energy $E_g$ only provides partial or noisy information on the update on $\mathbf{p}$, the optimization is usually accompanied with a regularization on $\mathbf{p}$. 

Popular regularizations on $\mathbf{p}$ are based on enforcing the smoothness of the surface. Usually, an additional smoothness term is defined and added to the original energy $E_g$ \cite{bogo2017dynamic, sela2017unrestricted, wang2018pixel2mesh}. Recent work on inverse rendering proposed new regularization that is based on re-parameterization of $\mathbf{p}$ \cite{nicolet2021large}. However, most regularizations in previous surface reconstruction frameworks only focus on the smoothness of $\mathbf{p}$ and do not handle the density of template mesh vertices.

Vertex density is an important parameter to control for accurate reconstruction of a shape using a fixed number of vertices in the template mesh. Regions with high curvatures require more position samples to express shape details accurately. We observe under-sampling in complex structures prohibits accurate reconstructions in both inverse rendering and non-rigid surface registration.

To this end, our key idea is to design a mesh density adaptation energy that is applied along with smoothness regularization. Our adaptation energy is defined with respect to edge lengths around each vertex. We calculate desired edge lengths to enforce desired adaptive density at each step of the iterative process for optimizing $E_g$. The desired adaptive density is calculated using intermediate reconstruction results to push vertices towards high-curvature regions.
Our density adaptation greatly improves the reconstruction of shape details by resolving the under-sampling problem (\Fig{teaser}).

In summary, our main contributions are as follows:
\begin{itemize}
    \item We present a novel mesh density adaptation method 
    that improves the reconstruction of shape details 
    in the template-based surface reconstruction.
    \item We apply mesh density adaptation to inverse rendering and show improvements in the reconstruction accuracy of complex structures.
    \item We present a non-rigid surface registration method based on mesh density adaptation that achieves faithful reconstruction of complex shapes and meaningful correspondences.
\end{itemize}

\section{Related Work}

\subsection{Template-based Shape Reconstruction}

Template mesh fitting is a widely studied approach to reduce the ill-posedness of 3D reconstruction.
Recent approaches in 3D shape reconstruction use spherical template meshes for reconstructing genus-0 3D shapes. Using the spherical template, reconstruction of 3D objects is simplified to a sphere deformation problem. In deep-learning based 3D reconstruction, single-view 3D reconstruction \cite{kato2018neural, wang2018pixel2mesh, wen2019pixel2mesh++} or multi-view 3D reconstruction \cite{chen2022multi} via deformation of spherical template mesh has been explored.

Recent template-based iterative reconstruction methods introduce new mechanisms for high-quality reconstruction.
In terms of optimization technique,
modifications to the Adam optimizer \cite{kingma2015adam} have been proposed to ensure smoothness \cite{nicolet2021large} or rotation equivariance \cite{ling2022vectoradam}.

In terms of surface regularization, Hanocka \Etal\ \shortcite{hanocka2020point2mesh} used convolutional neural networks to parameterize template deformation and regularize reconstruction with self-priors. Self-priors do not explicitly enforce smoothness, but periodic remeshing recovers a watertight manifold mesh. Nicolet \Etal\ \shortcite{nicolet2021large} introduced diffusion re-parameterization to propagate sparse gradients from image silhouettes. The smoothness from the diffusion helps maintaining a smooth manifold mesh throughout the iterative optimization. However, both methods do not address vertex density, potentially leading to sparse vertices, especially near extruded regions like a bunny's ear. To mitigate sparse vertices, periodic remeshing, which changes initial meshing in the template, was employed.

Template-based approach has achieved notable success for 3D reconstruction of human face, body, and animal, in addition to their non-rigid registration \cite{paysan20093d, li2017learning,  dai2018non, amberg2007optimal, schneider2009fast,sela2017unrestricted, bogo2014faust, bogo2017dynamic, hirshberg2012coregistration, yang2015sparse, zuffi20173d}. Nonetheless, the control of vertex density during reconstruction has not been explicitly addressed in those approaches. The lack of control on vertex density may lead to missing details due to under-sampling with sparse vertices on complex structures.

Our method provides the control for vertex density on a mesh in a template-based reconstruction.
Our mesh density adaptation method is based on template mesh deformation and preserves the initial vertex connectivity.
So, our method can be easily incorporated into the iterative deformation update step in a template-based 3D reconstruction framework. We demonstrate the performance of our method in inverse rendering and non-rigid shape registration.

\subsection{Remeshing and Mesh Optimization}

Remeshing has been extensively studied in geometry processing. The goal of remeshing is to obtain a 3D mesh that possess better meshing quality while retaining the original shape. For a thorough review on remeshing in general, refer to \cite{khan2020surface, botsch2010polygon-remesh}.

Remeshing frameworks handle adaptive density needed for preserving fine details in the original shape via re-sampling or re-triangulation. Alliez \Etal\ \shortcite{alliez2003anisotropic} use curvature-based adaptive sampling of vertices for anisotropic remeshing. Chen \Etal\ \shortcite{chen2012isotropic} employ curvature-based vertex sampling and mesh optimization for isotropic remeshing. Dey and Ray \shortcite{dey2010polygonal} perform isotropic remeshing to provide improved mesh density. Jakob \Etal\ \shortcite{jakob2015instant} handles adaptive resolution in remeshing by integrating custom metrics based on mean curvature.

While remeshing is useful for processing a fixed shape to obtain a better quality mesh, it is not suited for template-based reconstruction where the template mesh is deformed toward the target shape without vertex connectivity change.
Nealen \Etal\ \shortcite{nealen2006laplacian} proposes a mesh optimization method to improve mesh quality while retaining the initial vertex connectivity.
This method could be used as a post-process of template-based reconstruction for better mesh quality, but the mesh density is not handled in the method.

\section{Mesh Density Adaptation}
\label{sec:density}

\begingroup
\setlength{\columnsep}{5pt}

Our goal is to control local vertex density in template-based 3D shape reconstruction. Previous literature has defined density on a surface as a density function \cite{du1999centroidal, chen2011efficient}, which requires global parameterization. An alternative definition could be based on triangle area: small triangles in a region induce high vertex density and vice versa. Modeling mesh density via triangle area does not require global parameterization, but triangle area cannot uniquely determine vertex distance, e.g., a long-thin skewed triangle can have the same area as a small equilateral triangle.

In template-based shape reconstruction, triangle shape and edge length have been controlled via remeshing \cite{lyu2020differentiable, nicolet2021large,luan2021unified, palfinger2022continuous} to improve reconstruction quality. For example, Palfinger \shortcite{palfinger2022continuous} applies  an adaptive remeshing step after each shape update to keep optimal edge lengths. However, remeshing breaks the initial meshing, which can be problematic in applications like non-rigid surface registration, where shared meshing between the reconstruction results is crucial for the results to be useful, e.g., for building shape keys for animation and transferring textures.

\begin{wrapfigure}[4]{r}{0.16\textwidth}
\vspace{-13pt}
\includegraphics[width=0.16\textwidth]{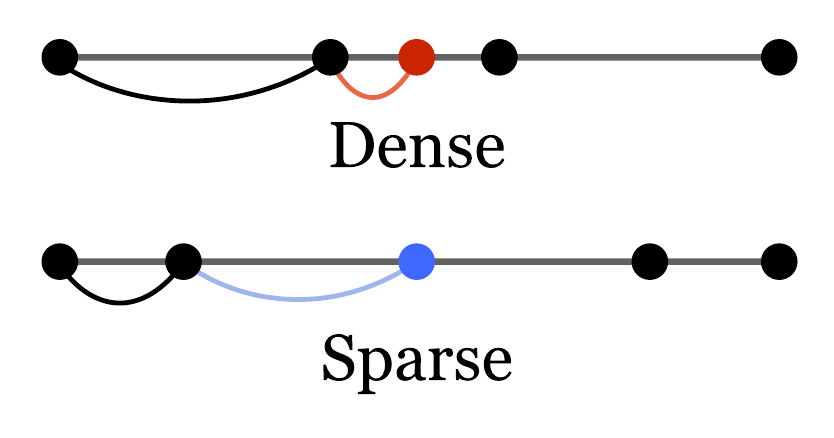}
\end{wrapfigure} 
In contrast, instead of remeshing, we use deformation to control mesh density. The deformation is guided by an adaptation energy based on edge length to enforce the desired mesh density. Edge lengths represent the distances between vertices. Intuitively, shorter edge lengths around a vertex result in denser sampling of vertices.

\endgroup

\subsection{Adaptation Energy}
We use edge lengths for adapting mesh density. 
Given an array of vertex positions $\mathbf{p} \in \mathbb{R}^{N \times 3}$, the average edge lengths $\mathbf{l}(\mathbf{p})$ for the vertices are calculated by
\begin{equation}
\mathbf{l}(p_i) = \dfrac{1}{|\mathcal{N}_i|}\sum_{j \in \mathcal{N}_i} \lVert {p}_i - {p}_j \rVert_2,
\end{equation} 
where $\mathbf{l}(p_i)$ is the average edge length for the i-th vertex $p_i$ and $\mathcal{N}_i$ is its 1-ring neighborhood. 

Using the edge lengths as a metric, we formulate the density-adaptation problem as optimizing the vertex position $\mathbf{p}$ to minimize the following adaptation energy $E_{a}$:
\begin{equation}
\begin{gathered}
\underset{\mathbf{p}}{\text{minimize  }} E_{a}(\mathbf{p}, \mathbf{l}'), \\
E_{a}(\mathbf{p}, \mathbf{l}') = \dfrac{1}{N} ||\mathbf{l}(\mathbf{p}) - \mathbf{l}' ||^2,
\end{gathered}
\label{eq:energy-adaptation}
\end{equation}
where $\mathbf{l}'$ denotes an array of desired edge lengths that encodes the target mesh density.
$N$ is the total number of vertices. 
Given the initial vertex position $\mathbf{p}_\text{init}$, we can iteratively update the vertex positions $\mathbf{p}$ by minimizing the energy $E_{a}$ via a gradient-based optimizer, such as Adam~\cite{kingma2015adam}.
In our template-based shape reconstruction, $\mathbf{p}_\text{init}$ is provided by the template mesh and $\mathbf{l}'$ is controlled to obtain a desirable mesh density of the reconstructed shape. Computation of desirable mesh density $\mathbf{l}'$ is detailed in \Sec{inverse}.

\subsection{Smoothness Regularization}

The solution of \Eq{energy-adaptation} may contain undesirable artifacts, such as self-intersections, depending on the target density $\mathbf{l}'$. We take \textit{diffusion re-parameterization}~\cite{nicolet2021large} as a regularization for \Eq{energy-adaptation} to avoid such problem.

Recently, Nicolet \Etal\ \shortcite{nicolet2021large} presented a modified gradient descent method for 3D mesh reconstruction from multi-view images. The modification effectively diffuses concentrated gradients from image silhouettes to other parts, allowing for stable deformation of a template mesh to target multi-view silhouettes. An insightful interpretation of their modified gradient descent method is that it is equivalent to ordinary gradient descent on the diffusion re-parameterization $\mathbf{u}$ defined as 
\begin{equation}
\mathbf{u}(\mathbf{p}) = (\mathbf{I} + \lambda \mathbf{L})\mathbf{p}, \label{eq:inverse-parameterization}
\end{equation}
\begin{equation}
\mathbf{p}(\mathbf{u}) = (\mathbf{I} + \lambda \mathbf{L})^{-1}\mathbf{u} \label{eq:re-parameterization}.
\end{equation}
$\mathbf{L}$ is the discrete Laplace operator \cite{taubin1995curve} and $\lambda$ is a constant weight.
The function $\mathbf{u}$ is diffused for a fixed time $\lambda$ using a backward Euler integration step \cite{baraff1998large}.

In this paper, we use the diffusion re-parameterization for mesh density adapation. In the gradient descent optimization, instead of directly updating the vertex position $\mathbf{p}$, we use gradients computed from \Eq{energy-adaptation} to update the re-parameterized coordinate $\mathbf{u}$ through \Eq{inverse-parameterization}, and 
the update of $\mathbf{p}$ is determined by the updated $\mathbf{u}$ through \Eq{re-parameterization}. Note that this indirect update of vertex position $\mathbf{p}$ differs from the direct update on $\mathbf{p}$, since \Eq{re-parameterization} defines $\mathbf{p}$ as diffused $\mathbf{u}$ and globally propagates the values of $\mathbf{u}$ onto $\mathbf{p}$.

Using diffusion re-parameterization, our final definition of the mesh density adaptation energy $E_{a}$ is given by
\begin{align}
&\underset{\mathbf{u}}{\text{minimize  }} E_{a}(\mathbf{p}(\mathbf{u}), \mathbf{l}'), 
    \label{eq:adaptation-re-parameterization}
\end{align} 
where the surface is regularized by simply changing the optimized variable from $\mathbf{p}$ to $\mathbf{u}$ in \Eq{energy-adaptation}.
We apply our adaptation energy in \Eq{adaptation-re-parameterization} to the tasks of inverse rendering (\Sec{inverse}) and non-rigid surface registration (\Sec{other}).

\paragraph{Laplacian regularization}
We briefly discuss the other alternative for smoothness regularization: the Laplacian regularization. We find Laplacian regularizer conflicts with the goal of our adaptation energy $E_a$. Common formulations for smoothness energy are \cite{sela2017unrestricted, wang2018pixel2mesh}:
\begin{equation}
\begin{gathered}
    E_{s}(\mathbf{p}) = \dfrac{1}{2}tr(\mathbf{p}^T\mathbf{L}\mathbf{p}) \text{ and} \\
    E_{s}(\mathbf{p}) = \dfrac{1}{2}\sum_i^N \lVert (\mathbf{L}\mathbf{p})_i \rVert^2 = \dfrac{1}{2}tr(\mathbf{p}^T\mathbf{L}^2\mathbf{p}),
    \label{eq:laplacian-regularizer}
\end{gathered}
\end{equation} where $tr(\cdot)$ is the trace of a matrix. 
The former is referred as the Laplacian energy and the latter as the bi-Laplacian energy.

In discrete surface meshes, the Laplace operator $\mathbf{L}$ is a sparse matrix $\mathbf{L} \in \mathbb{R}^{N \times N}$ that calculates the difference between each vertex's position and the weighted average of its 1-ring neighbors:
\begin{equation}
(\mathbf{L}\mathbf{p})_i = {p}_i - \sum_{j \in \mathcal{N}_i} w_{ij}{p}_j.
\label{eq:differential-coordinates-mesh}
\end{equation} 

Laplacian regularization forces a mesh density encoded in the discrete Laplace operator $\mathbf{L}$. For each $i$-th vertex, the Laplacian becomes small when ${p}_i$ coincides with the weighted average of the neighbors. Minimizing the Laplacian would move the vertices towards the weighted average of their neighbors, causing an adjustment in mesh density that may not match with our desired mesh density specified by $\mathbf{l}'$. Therefore, the Laplacian regularizers may conflict with our goal of adapting mesh density.

\section{Inverse Rendering}
\label{sec:inverse}

In this section, we apply our mesh density adaptation to the inverse rendering problem. 
Specifically, given a set of multi-view images, we reconstruct the 3D shape that matches the input images. We follow the experimental setup in Nicolet \Etal\ \shortcite{nicolet2021large}; we use a sphere as a template mesh and deform the sphere to a 3D shape that generates the target renderings at multiple views.

A fundamental challenge in this setup is the absence of correspondences between the template sphere and the input images. Then, approximated correspondences are used based on the intermediate registration results, and the correspondences are updated at each iteration.
Nicolet \Etal\ \shortcite{nicolet2021large} maintain smooth intermediate results during the updates using diffusion re-parameterization. The smoothness provided with diffusion re-parameterization helps obtain a rough global shape stably. However, because of no explicit constraint on the mesh density, results may suffer from loss of details due to under-sampling near complex structures.

In this paper, we improve inverse rendering by applying our mesh density adaptation. In the original article~\cite{nicolet2021large}, the under-sampling was resolved with periodic remeshing of intermediate results. Remeshing changes the mesh connectivity and increases the number of vertices in the template mesh.
In contrast, we propose a simple modification of the original energy to control mesh density and resolve the under-sampling problem without remeshing. Remeshing incurs recomputation of $(\mathbf{I} + \lambda \mathbf{L})^{-1}$ in \Eq{re-parameterization} and a change in the number of optimized variables. By retaining the initial mesh, inverse rendering benefits from a lower computational burden and a simpler formulation.

We briefly overview the framework for inverse rendering in \cite{nicolet2021large}. The shape is optimized by minimizing a loss that measures the discrepancy between the observed images and the rendered images using the 3D shape $\mathbf{p}$. Given a set of rendered images $\mathcal{I}$ with the corresponding camera poses $\mathcal{P}$ and the differentiable rendering function $\mathcal{R}$ \cite{laine2020modular}, the photometric loss $E_{p}$ is defined by
\begin{equation}
    E_{p}(\mathbf{p}, \mathcal{I}) = \sum_i \lVert \mathcal{R}(\mathbf{p}, \mathcal{P}_i) - \mathcal{I}_i \rVert_1.
\end{equation} 
The loss is minimized by optimizing the diffusion re-parameterization $\mathbf{u}$,
and the optimization problem is formulated as 
\begin{equation}
    \underset{\mathbf{u}}{\text{minimize }} E_{p}(\mathbf{p}(\mathbf{u}), \mathcal{I}).
    \label{eq:inverse-original}
\end{equation}

\subsection{Density Adaptation}
\label{sec:inverse-density-adaptaion}
We incorporate our adaptation energy $E_a$ in \Eq{adaptation-re-parameterization} into the original formulation in \Eq{inverse-original}. We add two terms with different goals in the density adaptation. One term is for enforcing the uniform density and the other is for enforcing adaptive high density near complex structures. Using intermediate shape $\mathbf{p}'$ obtained at the previous iteration in the optimization process, new optimization problem is
\begin{equation}
    \underset{\mathbf{u}}{\text{minimize }}  E_{p}(\mathbf{p}(\mathbf{u}), \mathcal{I}) + w_u E_{a}(\mathbf{p}(\mathbf{u}), \mathbf{l}'_u(\mathbf{p}'))+ w_k E_{a}(\mathbf{p}(\mathbf{u}), \mathbf{l}'_k(\mathbf{p}')),
    \label{eq:inverse-new}
\end{equation} 
where $\mathbf{l}'_u(\mathbf{p}')$ and $\mathbf{l}'_k(\mathbf{p}')$ encode the uniform and adaptive mesh densities, respectively. 
Each of $\mathbf{l}'_u(\mathbf{p}')$ and $\mathbf{l}'_k(\mathbf{p}')$ contains the desired mean edge length per vertex that is calculated using $\mathbf{p}'$.
$w_u$ and $w_k$ are the weights for the adaptation energy. By increasing $w_u$, the shape $\mathbf{p}$ computed at the current iteration would possess more uniform mesh density. By increasing $w_k$, the shape $\mathbf{p}$ would have higher densities around detailed structures. The control and the role of the two weights are discussed in detail in \Sec{scheduling}.

We design $\mathbf{l}'_u(\mathbf{p}')$ to enforce uniform mesh density. That is, all desired mean edge lengths for vertices are the same:
\begin{equation}
    \mathbf{l}'_u(\mathbf{p}') = l_m(\mathbf{p}'),
\end{equation} 
where $l_m(\mathbf{p}')$ is the average of mean edge lengths of vertices in $\mathbf{p}'$.

We design $\mathbf{l}'_k$ to increase the mesh density in the regions with complex structures that require relatively large numbers of vertices for accurate reconstruction.
We measure the structure complexity using the magnitude of Laplacian computed from $\mathbf{p}'$. Our assumption is that complex structures exhibit high mean curvatures, which is proportional to the magnitude of the Laplacian \cite{botsch2010polygon-curvature}. For a more detailed discussion, refer to \Sec{discussion-curvature}.

We first calculate the norm of Laplacians $\mathbf{K} \in \mathbb{R}^{N}$ from the intermediate shape $\mathbf{p}'$. The values are scaled with the average of mean edge lengths of vertices:
\begin{equation}
    K_i = \dfrac{1}{l_m(\mathbf{p}')}\lVert(\mathbf{L}\mathbf{p}')_i\rVert_2,
\end{equation} 
where the discrete Laplace operator $\mathbf{L}$ uses the uniform weight.
We then smooth out $\mathbf{K}$ over the surface to reduce noise using diffusion with one step of the backward Euler's method \cite{baraff1998large}. 
The smoothed value $\mathbf{S} \in \mathbb{R}^{N}$ is obtained by
\begin{equation}
    \mathbf{S} = (\mathbf{I} + \lambda_s\mathbf{L})^{-1}\mathbf{K},
\end{equation} 
where $\lambda_s$ is the time step that determines the degree of diffusion. We use $\lambda_s = 1$. 
We define the desired mean edge length $\mathbf{l}'_k$ by scaling the current mean edge length $\mathbf{l}$ using $\mathbf{S}$:
\begin{equation}
    \mathbf{l}'_k(\mathbf{p}') = \mathbf{l}(\mathbf{p}') \odot \textit{clamp}\left(\bar{S}\oslash\mathbf{S}\right),
    \label{eq:density-curvature}
\end{equation} 
where $\odot$ is element-wise multiplication, $\oslash$ is element-wise division, $\bar{S}$ is the average of scalar values in $\mathbf{S}$, 
and $\textit{clamp}(\cdot)$
clips the values in a vector to $[0,1]$.
As a result, the target mean edge lengths in $\mathbf{l}'_k$ are scaled down from the current values in $\mathbf{l}$ for the vertices whose Laplacian magnitudes are larger than the average. For other vertices, the target mean edge lengths in $\mathbf{l}'_k$ are the same as the values in $\mathbf{l}$.

\setcounter{figure}{1}\begin{figure}
    \centering
    \includegraphics[width=0.45\textwidth]{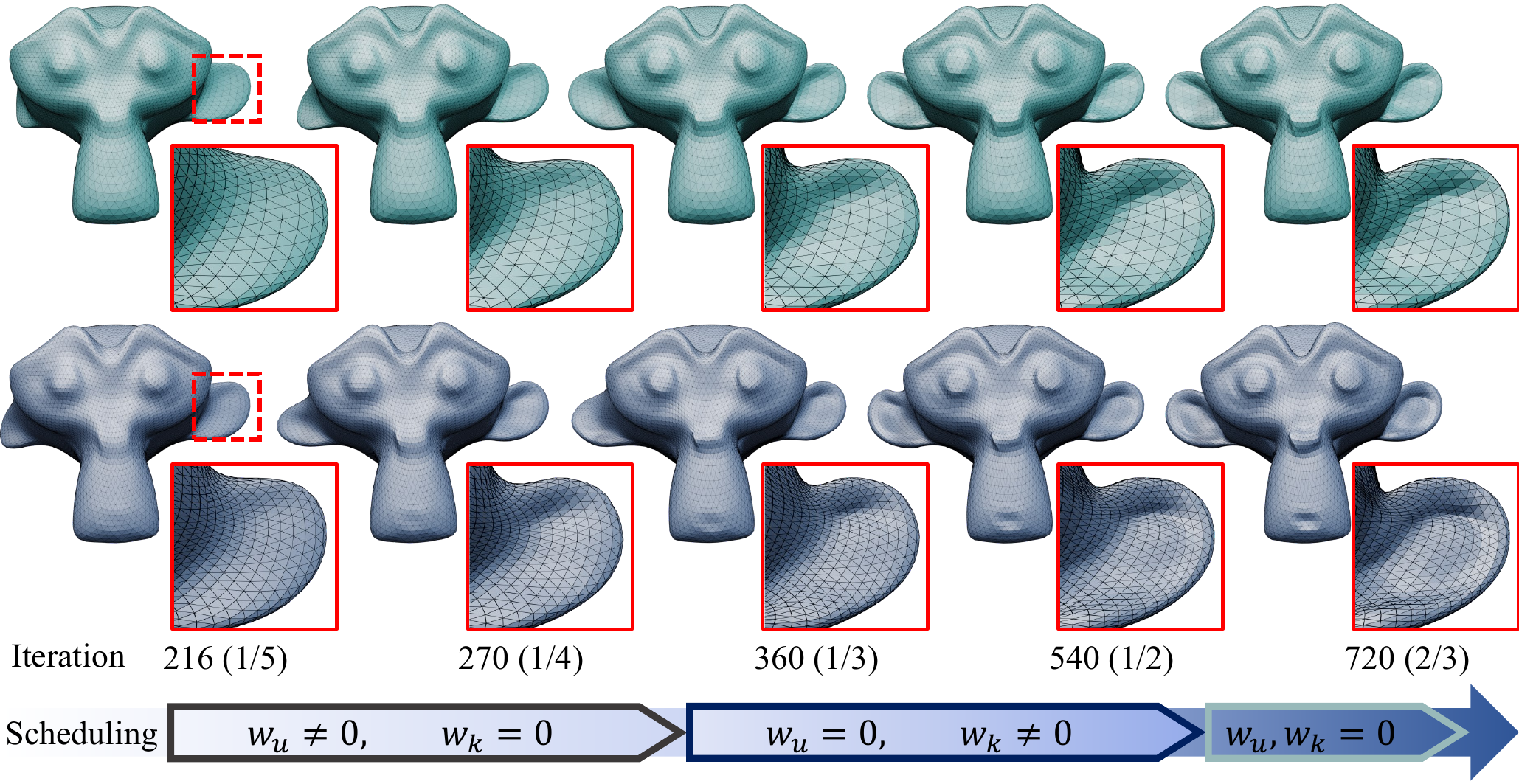}
    
    \caption{
    Visualization of our density adaptation and weight scheduling in inverse rendering. Top: Without our density adaptation. Bottom: With our density adaptation. We enforce uniform density until $1/4$ of the iterations, then use adaptive density to push vertices to high-curvature regions. The last half of the iterations are done without our adaptation energy. See the supplementary video\protect\footnotemark{} for more results.
    }
    
    \label{fig:vis-mean-area}
\end{figure}

\subsection{Weight Scheduling}
\label{sec:scheduling}

In the iterative optimization process for inverse rendering, we schedule the two weights $w_u$ and $w_k$ of the adaptation energy in \Eq{inverse-new}.
We obtain the rough global shape by using uniform mesh density in the early stages and recover shape details by using adaptive mesh density in the later stages.

In the first quarter of iterations, we use non-zero constant $w_u$ and zero $w_k$ to enforce uniform mesh density during the early development of shape from the sphere. In this early stage, the sphere is deformed to express a rough outline of the target shape,
and $\mathbf{p}'$ does not possess much meaningful shape information for inferring the adaptive mesh density. So, we prevent unnecessary adjustment in mesh density by enforcing uniform density. 
In the second quarter, we use zero $w_u$ and non-zero constant $w_k$ to apply density adaptation to push vertices towards complex structures. Finally, in the last half of iterations, we set both $w_u$ and $w_k$ to zero to focus on lowering the data loss $E_p$ in \Eq{inverse-new}. The effect of the weight scheduling is visualized in \Fig{vis-mean-area}.
    
\footnotetext{\url{https://youtu.be/L-WNBUNyP-Y}}

\subsection{Results}
\label{sec:inverse-rendering-results}

\begin{figure*}
    \begin{minipage}{0.12\textwidth}
    Laplacian
    \end{minipage}
    \begin{minipage}{0.13\textwidth}
    \ \ \ \ \ \ \ \ Bi-Laplacian
    \end{minipage}
    \begin{minipage}{0.2\textwidth}
    \ \ \ \ \ \ \ \ \ \ \ \cite{nicolet2021large}
    \end{minipage}
    \begin{minipage}{0.17\textwidth}
    \ \ \ \ \ \ \ \ \ \ \ \ \ \ Ours
    \end{minipage}
    \begin{minipage}{0.11\textwidth}
    \ \ \ \ \ \ \ \ \ \ \ \ \ GT
    \end{minipage}
    \centering
    \includegraphics[width=0.85\textwidth]{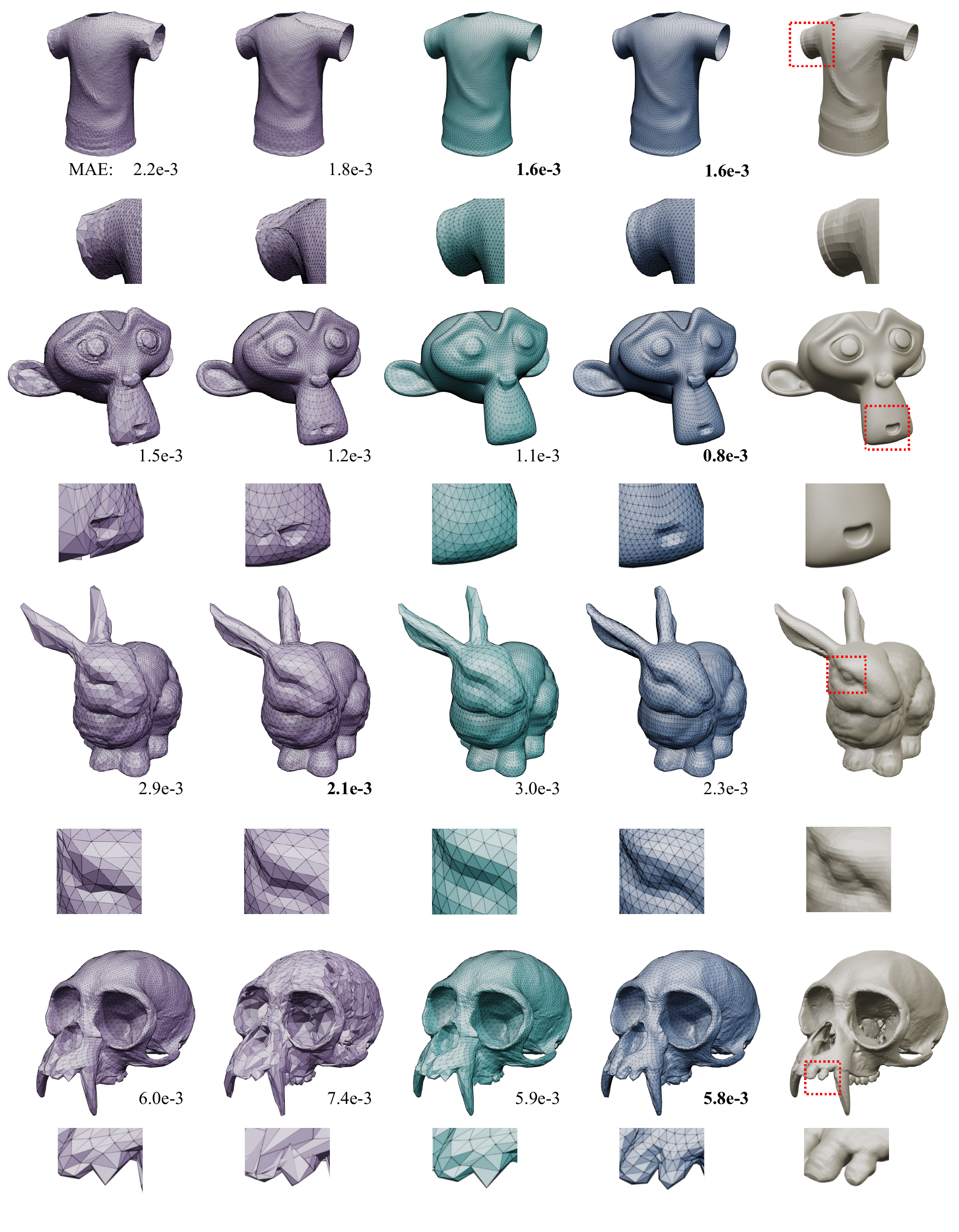}
    \caption{Comparison of different regularization methods in inverse rendering. Our method improves reconstruction of complex structures by assigning more vertices around detailed shapes. MAE refers to the mean absolute pixel error across all viewpoints. Note that MAE reported here is calculated from rendered images, while the errors in \Tbl{inverse-rendering-error} is calculated using the ground-truth mesh.}
    \label{fig:inverse-rendering}
\end{figure*}

In \Fig{inverse-rendering}, we compare our results with Laplacian regularizations and \cite{nicolet2021large}.
Regularizations with Laplacian and Bi-Laplacian energy terms are unstable due to the difficulty of balancing between reconstruction accuracy and surface smoothness.
For each scene, all methods use the same template mesh with limited vertex budget. We compare ours with the result of \cite{nicolet2021large} without remeshing.

Laplacian-regularized baselines use fine-tuned small regularization weights to focus on reconstruction accuracy, and the results exhibit noisy surfaces and irregular meshing. Results of \cite{nicolet2021large} show much smoother shapes, but they show loss of details due to under-sampling on complex structures and highly extruded regions. Our method assigns a proper amount of vertices to a region depending on the structure complexity and nicely recovers shape details, including highly extruded regions.

\Tbl{inverse-rendering-error} presents quantitative evaluation results for the T-shirt, Suzanne, Plank, Bob, and Bunny scenes from \cite{nicolet2021large}, measuring Chamfer distance and mean squared error of normal vectors between the ground-truth meshes and registration results. The Cranium scene is excluded because invisible surfaces beneath visible ones result in invalid Chamfer distances. We also report errors using only the top 5\% and 20\% salient vertices from ground-truth meshes, with saliency values calculated using \cite{lee2005mesh}. The results show that our density adaptation accurately reconstructs salient structures. For implementation details on the inverse rendering, refer to the supplementary document.

\begin{table}
\caption{
Comparison of reconstruction errors for the inverse rendering. In each cell, Left: Chamfer distance. Right: mean squared error of normal vectors. The averages of the five scenes are presented, along with the average errors for the top 20\% and 5\% salient vertices of the ground-truth meshes.
}
\label{tbl:inverse-rendering-error}
\resizebox{.65\linewidth}{!}{
\begin{tabular}{ccc}
    & \cite{nicolet2021large} &Ours \\
    \hline
    T-shirt         &0.0073/0.0107   &0.0073/0.0106   \\
    Suzanne         &0.0082/0.0153   &0.0080/0.0129   \\
    Plank           &0.0081/0.0071   &0.0081/0.0072   \\
    Bob             &0.0109/0.0061   &0.0107/0.0058   \\
    Bunny           &0.0092/0.0210   &0.0083/0.0167   \\
    \hline
    Average            &0.0087/0.0120   &\textbf{0.0085/0.0106}   \\
    top 20\% sal.   &0.0089/0.0219   &\textbf{0.0087/0.0205}   \\
    top 5\% sal.    &0.0093/0.0391   &\textbf{0.0089/0.0375}   \\
\end{tabular}
}
\end{table}

\section{Non-rigid Surface Registration}
\label{sec:other}

In this section, we demonstrate application of our mesh density adaptation to the non-rigid surface registration task. 
In our setting of non-rigid surface registration, given a set of 3D meshes, a template mesh is deformed to fit the input meshes while retaining the original template vertex connectivity. The resulting deformed template meshes should also share meaningful point-to-point correspondences.
Such shared point-to-point correspondences among multiple meshes with the same vertex connectivity enable useful applications, such as construction of probabilistic shape models \cite{paysan20093d, li2017learning}, texture transfer \cite{allen2003space}, and seamless shape blending \cite{yu2004mesh, sorkine2004laplacian}.

We find that accurate reconstruction of an input 3D mesh using a template mesh can be achieved by simple modification of our density-adapting inverse rendering framework introduced in \Sec{inverse}, where the only major difference would be the data loss to define the fitting. However, naive fitting of the template mesh may not provide meaningful point-to-point correspondences among the deformed template meshes, although the same vertex connectivity is guaranteed. A common approach for obtaining such correspondences is to annotate a set of sparse landmarks on the template \cite{amberg2007optimal} as well as on the input meshes. The known correspondences between the landmarks can guide the registration process to establish meaningful point-to-point correspondences among the template and the input meshes. To this end, we extend our density adaptation framework for template mesh fitting to reflect the correspondences of sparse landmarks.

We use a sphere as the template mesh as in \Sec{inverse} and reconstruct input meshes using adaptive mesh densities. Then, we resample on the sphere the sparse landmarks annotated on input meshes while reflecting the adaptive mesh densities of the reconstructed meshes. Finally, we fit the sphere again to input meshes allowing adaptive mesh densities, but now reflecting the correspondences among the resampled landmarks on the sphere and the landmarks on input meshes. In the following, we explain each step and show improvements in non-rigid template registration results using 3DCaricshop dataset \cite{qiu20213dcaricshop}, which contains artist-sculpted and exaggerated 3D faces.

\subsection{Template Mesh Fitting}
\label{sec:sphere-fitting}

Given a target 3D mesh $\mathcal{M}$ and the closest points $\hat{\mathbf{p}} \in \mathbb{R}^{N \times 3}$ on the surface $\mathcal{M}$ searched with intermediate result $\mathbf{p}$, the data loss for the fitting $E_d$ is defined as
\begin{equation}
E_d(\mathbf{p}, \hat{\mathbf{p}}) = D_c(\mathbf{p}, \hat{\mathbf{p}}) + D_n(\mathbf{p}, \hat{\mathbf{p}}),
\end{equation} where the Chamfer distance $D_c$ is
\begin{equation}
    D_c(\mathbf{p}, \hat{\mathbf{p}}) = \dfrac{1}{N} \sum_i^N \lVert p_i - \hat{p}_i \rVert_2
\end{equation} and the normal loss $D_n$ is
\begin{equation}
    D_n(\mathbf{p}, \hat{\mathbf{p}}) = \dfrac{1}{N} \sum_i^N \left(1 - n_i \cdot \hat{n}_i \right)
\end{equation} 
with vertex normals $n_i$ and $\hat{n}_i$ of $p_i$ and $\hat{p}_i$, respectively.

Replacing the photometric loss $E_p$ with $E_d$ in \Eq{inverse-new}, the optimization for template fitting is
\begin{equation}
    \underset{\mathbf{u}}{\text{minimize }} E_{d}(\mathbf{p}(\mathbf{u}), \hat{\mathbf{p}}) + w_u E_{a}(\mathbf{p}(\mathbf{u}), \mathbf{l}'_u(\mathbf{p}))+ w_k E_{a}(\mathbf{p}(\mathbf{u}), \mathbf{l}'_k(\mathbf{p})).
    \label{eq:nonrigid-base}
\end{equation}

\begin{figure*}
    \centering
    \begin{minipage}{0.19\textwidth}
    DR, 2.5K
    \end{minipage}
    \begin{minipage}{0.19\textwidth}
    Ours, 2.5K
    \end{minipage}
    \begin{minipage}{0.19\textwidth}
    \ \ \ DR, 10K 
    \end{minipage}
    \begin{minipage}{0.15\textwidth}
    \ \ \ Ours, 10K
    \end{minipage}
    \begin{minipage}{0.15\textwidth}
    \ \ \ \ \ \ \ \ \ \ \ \ \ \ \ \ \ \ GT
    \end{minipage}
    \includegraphics[width=0.99\textwidth]{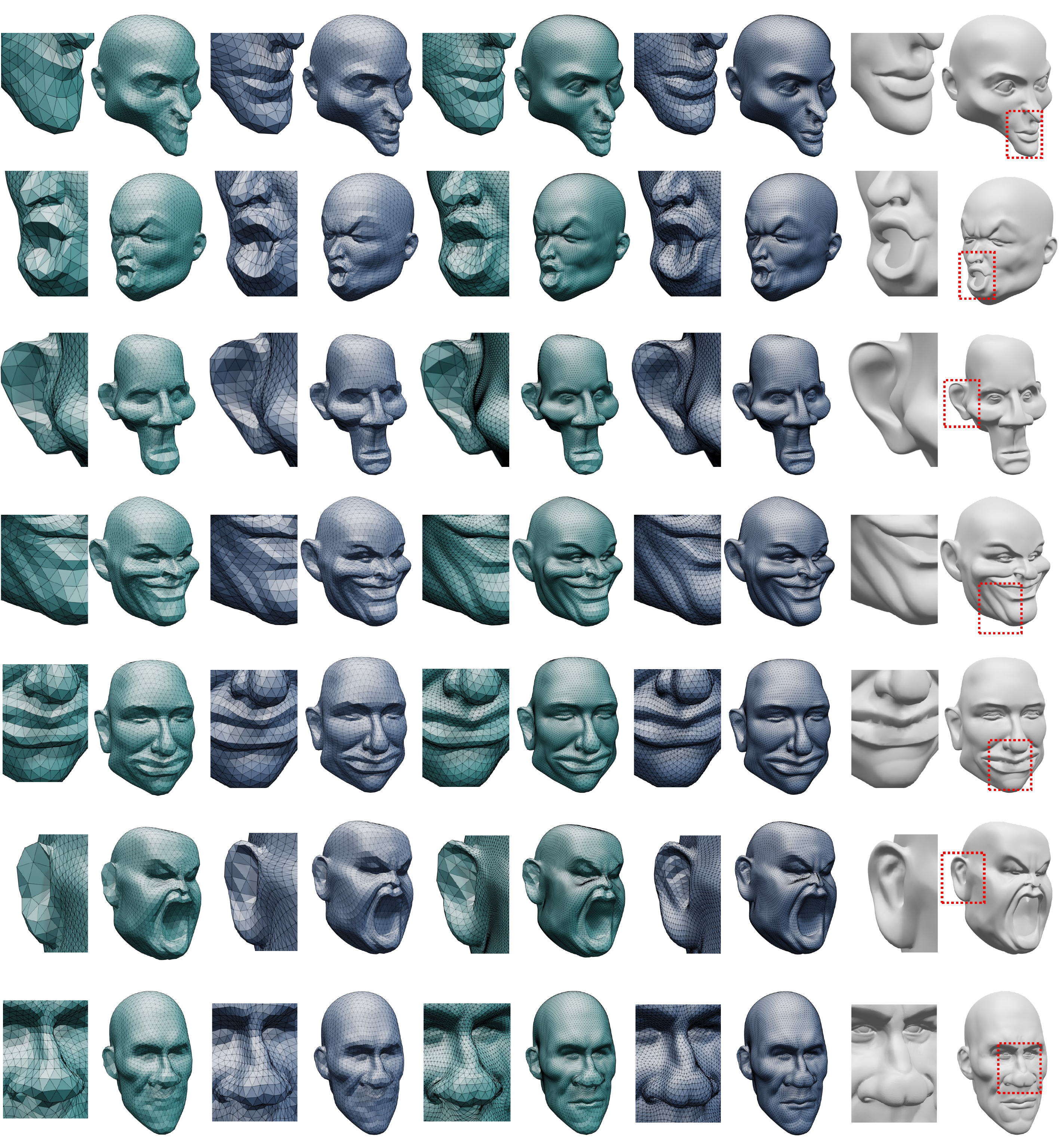}\\
    \caption{Template sphere fitting results to artist-sculpted cartoon meshes in \cite{qiu20213dcaricshop}. Our mesh density adaptation enables accurate reconstruction of the ground-truth shapes around complex structures. DR refers to only applying diffusion re-paramterization without our mesh density adaptation. Two results per method are shown: using 2.5K vertices and 10K vertices for the template sphere.}
    \label{fig:sphere-fitting}
\end{figure*}

The results of \Eq{nonrigid-base} are presented in \Fig{sphere-fitting}. Artist-sculpted 3D cartoon faces in 3DCaricShop \cite{qiu20213dcaricshop} are registered using our method. We compare our results with the case of not applying our mesh density adaptation (i.e., $w_u = w_k = 0$ in \Eq{nonrigid-base}), where the shape is only regularized with diffusion re-parameterization. We test the two methods in two settings for the template sphere mesh: with 2.5K vertices and 10K vertices. In both cases, our density adaptation enables accurate reconstruction of details by placing more vertices around complex structures.

\subsection{Landmark Resampling}
\label{sec:template-generation}

We aggregate input mesh landmarks to construct template sphere landmarks.
Given a set of input meshes with landmark annotations, we first fit the template sphere to the input meshes using the optimization in \Eq{nonrigid-base}. For each fitting $\mathbf{p}_i$, we determine the landmark vertex indices on the template sphere by finding the closest vertex in $\mathbf{p}_i$ for each landmark in the input mesh $\hat{\mathbf{p}}_i$. 
We use $\mathbf{c}_i$ to denote the landmark positions on the sphere corresponding to the landmark vertex indices from the $i$-th fitting $\mathbf{p}_i$.

We then select $\mathbf{c}_1$ as the reference set of landmarks and align other landmark sets to $\mathbf{c}_1$ by computing a rotation matrix $\mathbf{R}_i$ for each $\mathbf{c}_i$.
The rotation matrix is calculated \cite{sorkine2009least} as 
\begin{equation}
    \mathbf{R}_i = \mathbf{U}_i\mathbf{V}_i^T,
\end{equation} where orthogonal matrices $\mathbf{U}$ and $\mathbf{V}$ are computed using SVD
\begin{equation}
    \mathbf{c}_i^T \mathbf{W} \mathbf{c}_1  = \mathbf{U}_i\mathbf{\Sigma}_i \mathbf{V}_i^T.
\end{equation} 
$\mathbf{\Sigma}$ is a diagonal singular value matrix. $\mathbf{W}$ is a diagonal matrix for weighting the importance of landmarks. We set a high weight for the landmark on the tip of the nose. For more details on the landmark weighting, refer to the supplementary material.

Finally, the landmarks are selected as the vertices on the sphere mesh closest to $\bar{\mathbf{c}}$, which is the average of the aligned landmarks:
\begin{equation}
    \bar{\mathbf{c}} = \dfrac{1}{M}\sum_i^M \mathbf{c}_i \mathbf{R}_i^T,
\end{equation} where $M$ is the number of input meshes.

\subsection{Non-rigid Registration with Landmarks}
\label{sec:registration-landmark}

Using the resampled landmarks on the sphere, our non-rigid registration is formulated similarly to \Eq{nonrigid-base}, but with an additional landmark loss term:
\begin{equation}
    \underset{\mathbf{u}}{\text{minimize }} E + \dfrac{1}{B}\sum_i^B \lVert c_i - k_i \rVert^2,
\end{equation} where $E=E_d+w_uE_a+w_kE_a$ is the original loss in \Eq{nonrigid-base}, $c_i$ is a landmark position on $\mathbf{p}(\mathbf{u})$ during the optimization, $k_i$ is the corresponding landmark on the target 3D shape example, and $B$ is the number of landmarks.

\begin{figure}
    \centering
    \begin{tabular}{cc}
    \rotatebox{90}{\ \ \ \ \ \ \ \ \ \ \ GT\ \ \ \ \ \ \ \ \ \ \ \ \ \ \ \ \ \ Ours \ \ \ \ \ \ \cite{qiu20213dcaricshop}}
    & 
    \includegraphics[width=0.36\textwidth]{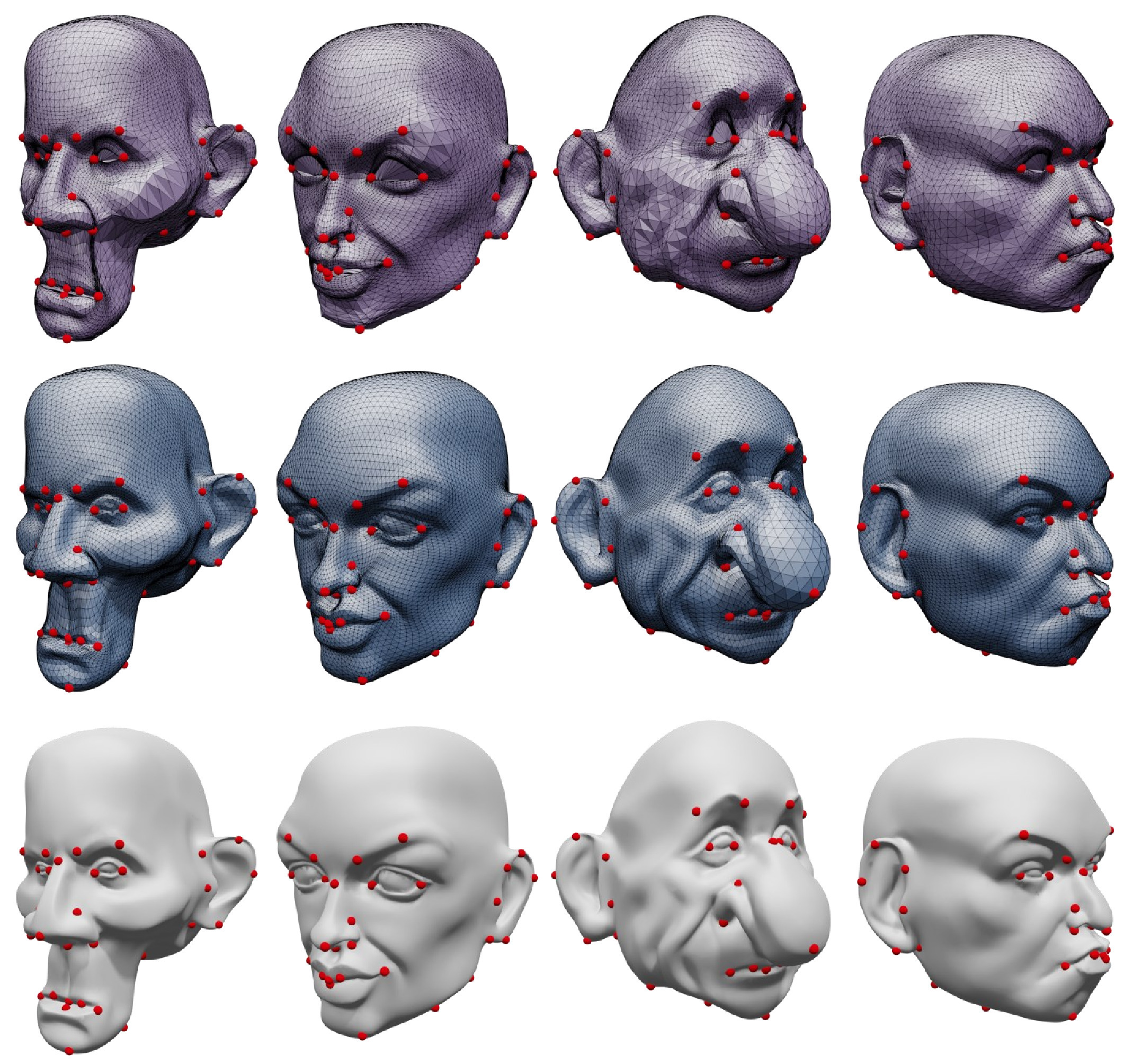}
    \end{tabular}
    \caption{
    Visual results of our proposed non-rigid registration. Landmark vertices are marked in red. Our method controls mesh density in both template construction and registration, resulting in 
    visual improvements over \cite{qiu20213dcaricshop}
    even with a fewer number of vertices in the template mesh. See the supplementary document for more results.
    }

    \label{fig:caricshop}
\end{figure}

\subsection{Results}
\label{sec:nonrigid-result}

We evaluate our non-rigid registration method on 3DCaricShop \cite{qiu20213dcaricshop} dataset, which contains artist-sculpted cartoon 3D models with highly exaggerated facial features. The dataset provides two sets of meshes: Raw meshes and registered meshes. The raw meshes were sculpted by artists and are highly detailed, but do not share vertex connectivity. The registered meshes provided by the authors are the results of running NICP \cite{amberg2007optimal} on the raw meshes using a 3D face template with 11,551 vertices.

In \Fig{caricshop}, we compare the results of our proposed non-rigid registration and the original NICP results provided with the dataset. NICP is the standard surface registration method in 3D face domain and used for constructing 3D face datasets, such as BFM \cite{paysan20093d} and LSFM \cite{booth2018large}. Our template mesh is a landmark-annotated sphere mesh with 10K vertices. Our method achieves detailed and accurate reconstruction compared to the original registration, using even lower vertex budget. %

The key factor behind our accurate registration is that we do not use a manually authored template mesh; We use a sphere as the template and determine the landmarks on the sphere in a data-driven manner. 
Although the meshing of our sphere template mesh is uniform, the spacing between landmarks on the sphere reflects the shape complexity of the region between the landmarks in 3D shape examples. Note that landmark spacing on the sphere is determined using the fitting results of 3D shape examples with adaptive density control, and the spacing increases for highly detailed regions.

\begin{table}
\caption{
Comparison of errors for non-rigid reconstruction. Left: Chamfer distance. Right: MSE of normal vectors. DR is our method without density adaptation. \textit{lmk} is landmark usage. Numbers in parenthesis show template vertex count.
}
\label{tbl:error-nonrigid}
\resizebox{.995\linewidth}{!}{
\begin{tabular}{cccccc}
    &\cite{qiu20213dcaricshop} (11K)&DR (2.5K)&Ours (2.5K)&DR (10K)&Ours (10K)\\
    \hline
    w/o lmk &     -/-           & 8.6e-5/0.025 & 8.7e-5/0.024 & 7.4e-5/0.016  & 6.9e-5/0.013\\
    top 20\% sal. &     -/-     & 1.6e-4/0.085 & 1.5e-4/0.078 & 1.3e-4/0.056  & 1.0e-4/0.045\\
    top 5\% sal. &     -/-      & 2.8e-4/0.115 & 2.5e-4/0.111 & 2.3e-4/0.084  & 1.7e-4/0.075\\
    \hline
    w/ lmk        &   0.046/0.11     & 8.5e-5/0.025 & 8.6e-5/0.025 & 7.0e-5/0.014 & \textbf{6.8e-5/0.013}\\
    top 20\% sal. &   0.022/0.22     & 1.4e-4/0.082 & 1.4e-4/0.081 & 1.1e-4/0.049 & \textbf{1.0e-4/0.046}\\
    top 5\% sal.  &   0.022/0.27     & 2.2e-4/0.113 & 2.2e-4/0.114 & 1.7e-4/0.079 & \textbf{1.6e-4/0.076}\\

\end{tabular}
}
\end{table}

\begin{figure}
    \centering
    \includegraphics[width=0.38\textwidth]{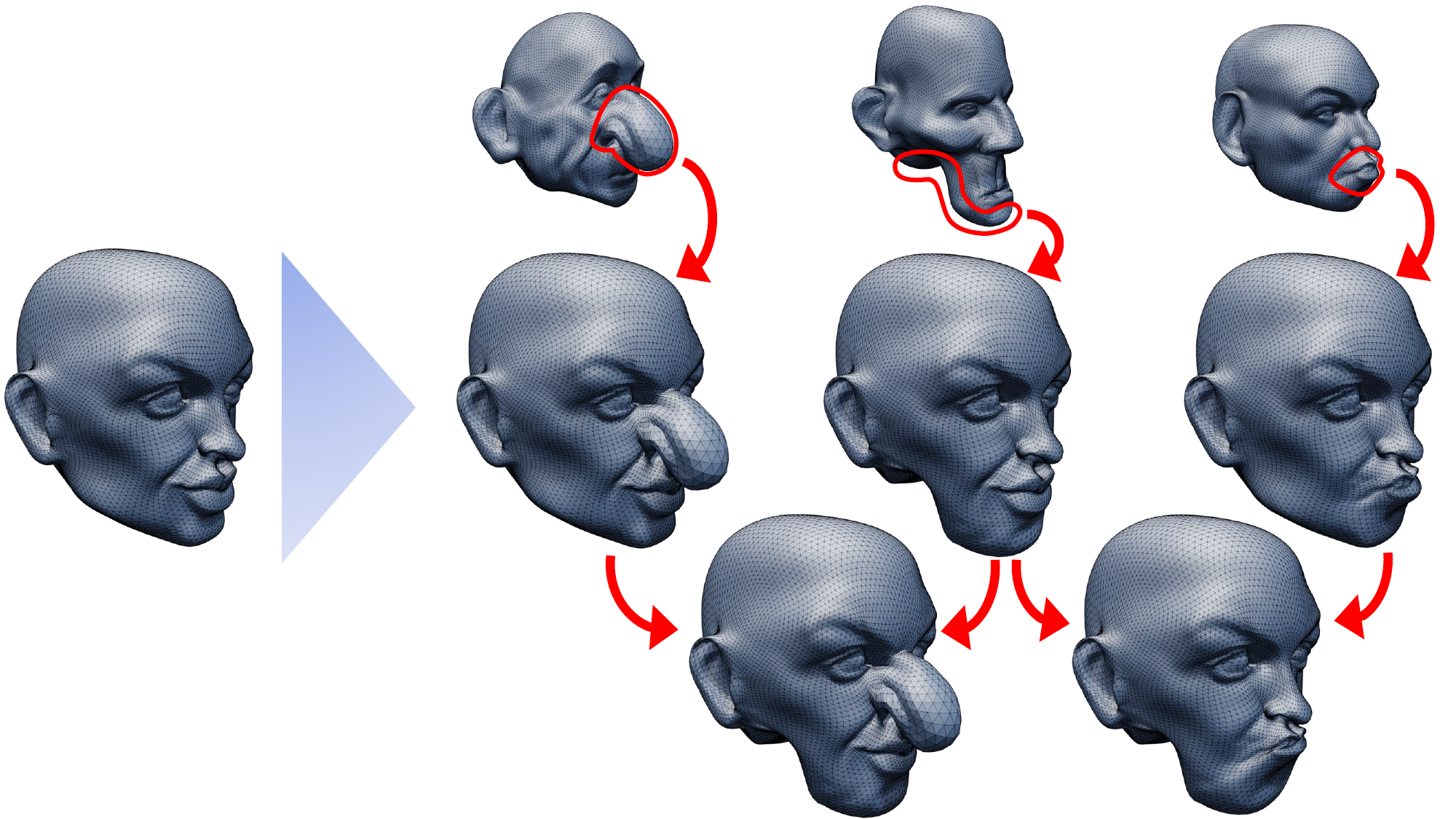}
    \caption{
    Seamless shape transplanting using dense correspondences from our non-rigid registration, with Laplacian-based transplanting \cite{sorkine2004laplacian} applied to 3D cartoon faces. Red curves mark the transplanted regions. Transplanting has been possible due to correct correspondences.
    }
    
    \label{fig:trans_plant}
\end{figure}

\Tbl{error-nonrigid} shows quantitative evaluation on our non-rigid registration. We use the same error metric as in \Sec{inverse-rendering-results}. We perform fitting on hand-sculpted meshes of the 3DCaricShop dataset \cite{qiu20213dcaricshop} and measure average errors after aligning landmarks with similarity transformation. Original fittings in 3DCaricShop show the highest error due to sparse vertices around facial components. Surface registration based on diffusion re-parameterization without density adaptation reduces the error significantly, but suffers from sub-optimal vertex distribution to express details in the target shape. Our method with density adaptation achieves the lowest error by distributing vertices effectively to express the details.

Our density adaptation is effective when a sufficient number of vertices is given; While the results with 2.5K template vertices do not show large improvements in term of the error metric, using 10K template vertices results in a significant drop in the error compared to the case of not using the density adaptation.

Using our re-sampled landmarks in the registration lowers the error while additionally granting dense correspondence between the shapes. \Fig{trans_plant} shows that the dense point-to-point correspondence obtained from our non-rigid registration is valid by applying it for shape transplanting \cite{sorkine2004laplacian}. Due to the dense correspondence, the Laplacian computed from a region in the source shape can be copied onto the corresponding region in the base mesh. Dense correspondence also enables texture transfer, as shown in the supplementary document.

\section{Discussion}
\label{sec:discussion-curvature}

Diffusion re-parameterization \cite{nicolet2021large} has been used for smooth propagation of sparse silhouette gradients in inverse rendering.
In this paper, we utilize diffusion re-parameterization for smoothing high-frequencies in dense gradients from density adaptation energy computed at all vertices.
Simply applying the update using the gradients from the density adaptation term without diffusion re-parameterization may lead to a noisy surface.

Our algorithm is effective at assigning appropriately high vertex density in large extruded regions, such as long ears of a monkey or bunny. Large extruded regions are manifested early in the iterative optimization, so the curvature changes in those regions are easily captured. Refer to the supplementary video for the visualization of the progression of the optimization.

\paragraph{Limitations}
Ideally, in addition to extruded structures, fine details would have benefited from high vertex densities in shape reconstruction. However, our current method cannot attract vertices to fine-scale details on wide and flat regions because fine details do not manifest until the late stage in the optimization. Visualization of this limitation is in the supplementary document. Handling fine structures in the density adaptation is left as future work.

\section{Conclusion}

We presented a novel mesh density adaptation method that improves the results of template-based surface reconstruction. Our method provides the control on vertex density of a mesh using a simple regularization loss based on mean edge length per vertex. The regularization is easily incorporated into existing optimization-based reconstruction tasks, such as inverse rendering and non-rigid surface registration. Our approach fills an often overlooked missing piece in the regularization, mesh density control, for accurate surface reconstruction via template deformation.

\begin{acks}

We thank anonymous reviewers for valuable feedback. This work was supported by IITP grants (AI Innovation Hub, 2021-0-02068;  AI Graduate School Program (POSTECH), 2019-0-01906), NRF grant (RS-2023-00211658), KOCCA grant (R2021040136), and KEIT grant (Alchemist Project, 20025752) from Korea Government (MSIT, MCST, and MOTIE) and Samsung Electronics.

\end{acks}

\newpage
\bibliographystyle{ACM-Reference-Format}
\bibliography{main}

\end{document}

\newcommand{\Eq}[1]  {Eq.~(\ref{eq:#1})}
\newcommand{\Eqs}[1] {Eqs.~(\ref{eq:#1})}
\newcommand{\Fig}[1] {Fig.~\ref{fig:#1}}
\newcommand{\Figs}[1]{Figs.~\ref{fig:#1}}
\newcommand{\Tbl}[1]  {Table~\ref{tbl:#1}}
\newcommand{\Tbls}[1] {Tables~\ref{tbl:#1}}
\newcommand{\Sec}[1] {Sec.~\ref{sec:#1}}
\newcommand{\Secs}[1] {Secs.~\ref{sec:#1}}
\newcommand{\Etal}{{\textit{et~al.}}}
\newcommand{\Alg}[1] {Algorithm~\ref{alg:#1}}

\newcommand{\setone}[1] {\left\{ #1 \right\}} %
\newcommand{\settwo}[2] {\left\{ #1 \mid #2 \right\}} %

\newcommand{\todo}[1]{{\textcolor{blue}{TODO: #1}}}
\newcommand{\change}[1]{{\color{red}#1}}
\newcommand{\sean}[1]{{\textcolor{magenta}{sean: #1}}}
\newcommand{\jung}[1]{{\textcolor{magenta}{jung: #1}}}
\newcommand{\kim}[1]{{\textcolor{magenta}{kim: #1}}}
\newcommand{\hwang}[1]{{\textcolor{magenta}{hwang: #1}}}
\newcommand{\baek}[1]{{\textcolor{magenta}{baek: #1}}}

\author{Yucheol Jung}
\authornote{Equal contribution}
\email{ycjung@postech.ac.kr}
\author{Hyomin Kim}
\authornotemark[1]
\email{min00001@postech.ac.kr}
\affiliation{%
 \institution{POSTECH}
 \city{Pohang}
 \country{South Korea}}

\author{Gyeongha Hwang}
\email{ghhwang@yu.ac.kr}
\affiliation{%
 \institution{Yeungnam University}
 \city{Gyeongsan}
 \country{South Korea}}

\author{Seung-Hwan Baek}
\email{shwbaek@postech.ac.kr}
\author{Seungyong Lee}
\email{leesy@postech.ac.kr}
\affiliation{%
 \institution{POSTECH}
 \city{Pohang}
 \country{South Korea}}
 
\makeatletter
\let\@authorsaddresses\@empty
\makeatother

\title[Mesh Density Adaptation for Template-based Shape Reconstruction]
    {- Supplementary Material -\\ Mesh Density Adaptation for \\ Template-based Shape Reconstruction}

\maketitle

In this supplementary document, we provide technical details, additional discussion, and more visual results. In \Sec{details}, we provide more information on the hyperparameters for the experiments conducted in the main paper.
In \Sec{smoothness}, we provide an additional experiment and discussion on the behavior of diffusion re-parameterization in the context of density adaptation.
In \Secs{fitting} and \ref{sec:registration}, we provide additional visual results for inverse rendering and non-rigid registration, respectively. \Sec{limitation} provides visual results for the limitation.

\section{Implementation Details}
\label{sec:details}

\subsection{Inverse Rendering}

Given adaptation strength $m$ and the number of total iterations $T$, the scheduling for the weights $w_u$ and $w_k$ of our adaptation energies at the $t$-th iteration is defined as
\begin{equation}
\begin{split}
w_u &= 
\begin{cases}
    m & \text{ if }\frac{t}{T} < \frac{1}{4} \\
    0 & \text{ otherwise} \\
\end{cases} \\
w_k &= 
\begin{cases}
    2m & \text{ if } \frac{1}{4} \leq \frac{t}{T} < \frac{1}{2} \\
    0 & \text{ otherwise}. \\
\end{cases}
\end{split}
\end{equation}
The adaptation strength per reconstructed object is presented in \Tbl{adaptation-strength}.

The weights for Laplacian regularizations differ per reconstructed object. We reuse the regularization weights used for the comparison in \cite{nicolet2021large}. We list the values for the Laplacian regularizations in \Tbl{regularization}.

For \cite{nicolet2021large} and ours, we use $\lambda=19$. Our method shares the same number of iterations as \cite{nicolet2021large} using the UniformAdam optimizer in \cite{nicolet2021large} with the same learning rate. Both \cite{nicolet2021large} and ours do not employ learning late decay, so small high-frequency shakes around the end of the optimization may occur (see the supplementary video). We list the number of total iterations and learning rate per object in \Tbl{optimization}. The regularization weights and learning rates per example in \Tbl{regularization} and \Tbl{optimization} are set to the same values as in \cite{nicolet2021large}.

\begin{table}[t]
    \caption{Adaptation strength used for \textit{Ours} in our experiments on inverse rendering}
    \label{tbl:adaptation-strength}
    \centering
    \begin{tabular}{c c c c c c}
        T-shirt & Suzanne & Bunny & Cranium & Plank & Bob \\
        \hline
        0.5 & 1.8 & 2.0 & 2.7 & 0.7 & 0.7
    \end{tabular}
\end{table}
\begin{table}[t]
    \caption{Regularization weights used for the compared Laplacian regularizations in our experiments on inverse rendering.}
    \label{tbl:regularization}
    \centering
    
\resizebox{.995\linewidth}{!}{
    \begin{tabular}{c c c c c c c}
        &T-shirt & Suzanne & Bunny & Cranium & Plank & Bob \\
        \hline
         Laplacian & 12 & 2.8 & 3.8 & 0.21 & 3.8 & 0.67\\
         Bi-Laplacian & 12 & 3.8 & 2.1 & 0.16 & 5.0 & 0.37
    \end{tabular}}
\end{table}

\subsection{Non-rigid Surface Registration}

\paragraph{Template mesh fitting}
For our result, we use the same weight scheduling for density adaptation as in the inverse rendering. We fix $m=1.5$ and the number of total iterations $T=1400$. For both \textit{Ours} and \textit{DR}, diffusion re-parameterization without density adaptation, we use $\lambda=19$.

\paragraph{Landmark resampling}
In the alignment of landmarks on sphere, we use a diagonal weight matrix
\begin{equation}
    \mathbf{W}_{ii} = \begin{cases}
        1.0\times10^4 & \text{ if } i = 16 \\
        1 & \text{ otherwise},
    \end{cases}
\end{equation} where $16$ is the index for the landmark on the tip of the nose. The total number of landmarks is $38$. The landmarks for the artist-sculpted meshes are obtained from 3DCaricShop \cite{qiu20213dcaricshop}. We used all faces in 3DCaricShop to perform the landmark resampling.

\begin{table*}[t]
    \centering
    \caption{Number of iterations and learning rate for the inverse rendering experiment.}
    \label{tbl:optimization}
    
    \resizebox{.7\linewidth}{!}{
    \begin{tabular}{c | c c c c c c}
        & T-shirt & Suzanne & Bunny & Cranium & Plank & Bob \\
        \hline
        \# input views & 16 & 13 & 16 & 16 & 16 & 16 \\
        \hline
        \# iter (Laplacian \& Bi-Laplacian) & 390 & 1130 & 1450 & 1910 & 960 & 940 \\
        \hline
        \# iter (Ours \& \cite{nicolet2021large}) & 370 & 1080 & 1380 & 5000 & 1500 & 930 \\
        \hline
        learning rate & 3E-3 & 2E-3 & 1E-2 & 5E-3 & 3e-3 & 3e-3\\
    \end{tabular}}
\end{table*}

\begin{figure*}[t]
    \centering
    \includegraphics[width=0.90\textwidth]{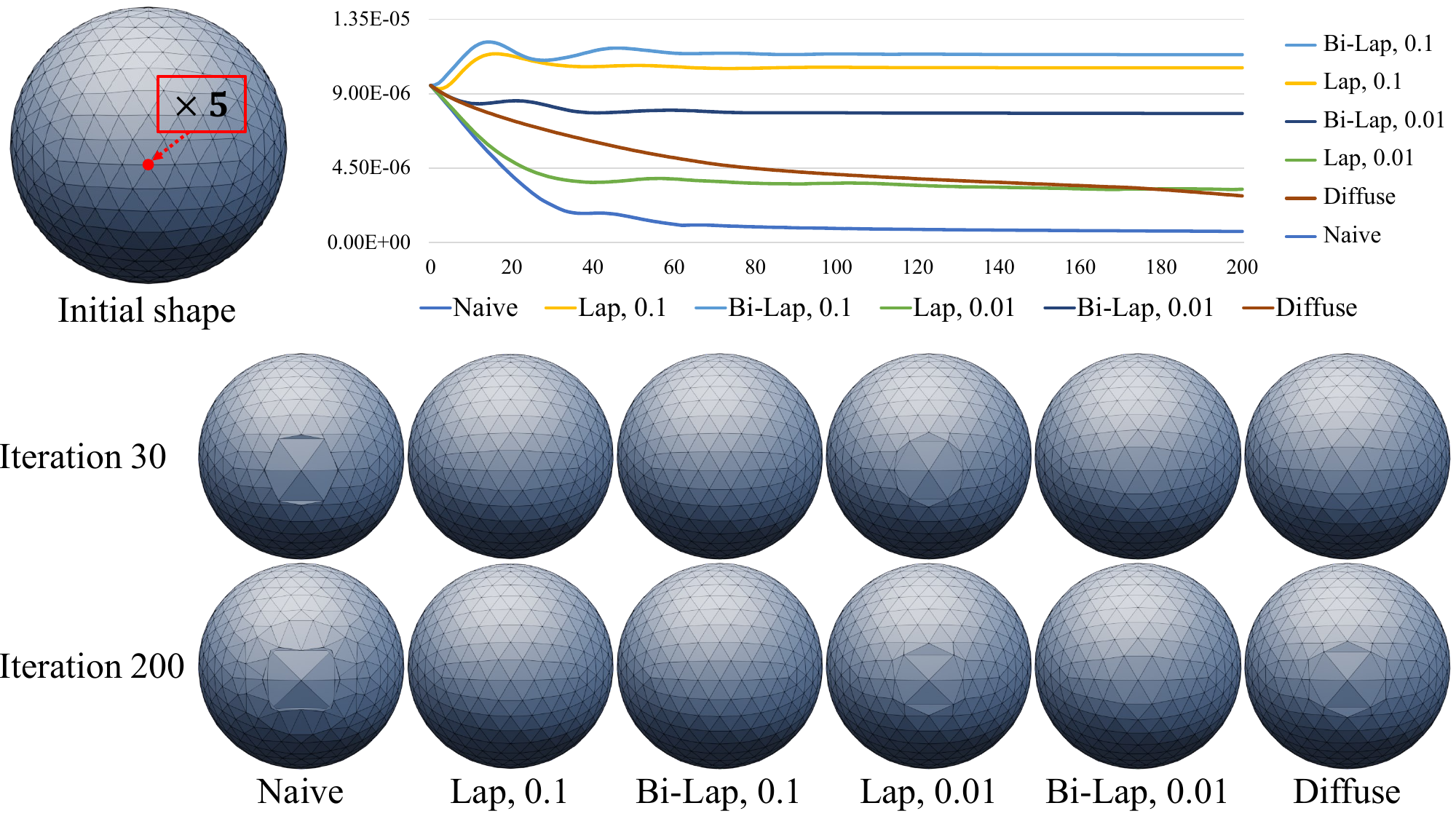}
\caption{
The behavior of different regularizations. We show the results of minimizing our adaptation energy with $\mathbf{l}'$ constructed by multiplying the mean edge length of one vertex by five. $\mathbf{p}$ is constrained to be on a sphere.
(Naive) Without any regularization, the update is instant, and the error is minimized in a short time.
(Lap, Bi-Lap) Results with Laplacian or Bi-Laplacian energy with two different weights are shown. These regularizations conflict with our goal of adapting the mesh density, so giving high weight ends up with non-adjustment of the density. Using a smaller weight, Lap\_0.01 allows the adaptation of mesh density but results in a quick update at almost same rate as the Naive case.
(Diffuse) \cite{nicolet2021large} does not conflict with density adaptation and prioritizes smooth deformation in the early stage, showing gradual decrease in the error and the eventual error minimization.
}

    \label{fig:comparison-regularization}
\end{figure*}

\subsection{Running time}
Our density adaptation term increases the optimization time by 10-15\%. In inverse rendering, \cite{nicolet2021large} takes 28 seconds and ours takes 32 seconds on  average to reconstruct the shape of one instance. In non-rigid registration, the optimization takes 19 seconds without density adaptation and takes 22 seconds with our density adaptation.

\section{Behavior of diffusion re-parameterization}
\label{sec:smoothness}

Diffusion re-parameteterization helps density adaptation by stabilizing the optimization. Our density adaptation is based on the examination of Laplacian of intermediate results. However, because Laplacian is sensitive to local shape change, the intermediate results may contain noisy Laplacian. So, the adjustment in the density based on the Laplacian should not be instant. Quick fluctuations from instant adjustment may induce unstable optimization.

Diffusion re-parameterization generally prefers smooth deformation but eventually allows adjustment of regularity if the gradients are accumulated enough. This slow-starting but eventually-achieving behavior is presented in \Fig{comparison-regularization}. In this sense, we find diffusion re-parameterization is an appropriate smoothness regularization for our method because it prevents quick fluctuations in the density adjustment while not conflicting with our goal.

\section{Additional Results for Inverse Rendering}
\label{sec:fitting}

\begin{figure*}[htbp]
    \begin{minipage}{0.32\textwidth}
    \ \ \ \ \ \ \ \ \ \ \ \ \  \ \ \ \ \ \ \ \ \ \cite{nicolet2021large}
    \end{minipage}
    \begin{minipage}{0.32\textwidth}
    \ \ \ \ \ \ \ \ \ \ \ \ \ \ \ \ \ \ \ \ \ \ \ \ \ \ \ \ \ \ \ \ \ Ours
    \end{minipage}
    \begin{minipage}{0.32\textwidth}
    \ \ \ \ \ \ \ \ \ \ \ \ \ \ \ \ \ \ \ \ \ \ \ \ \ \  \ \ \ \ \ \ GT
    \end{minipage}
    \centering
    \includegraphics[width=0.90\textwidth]{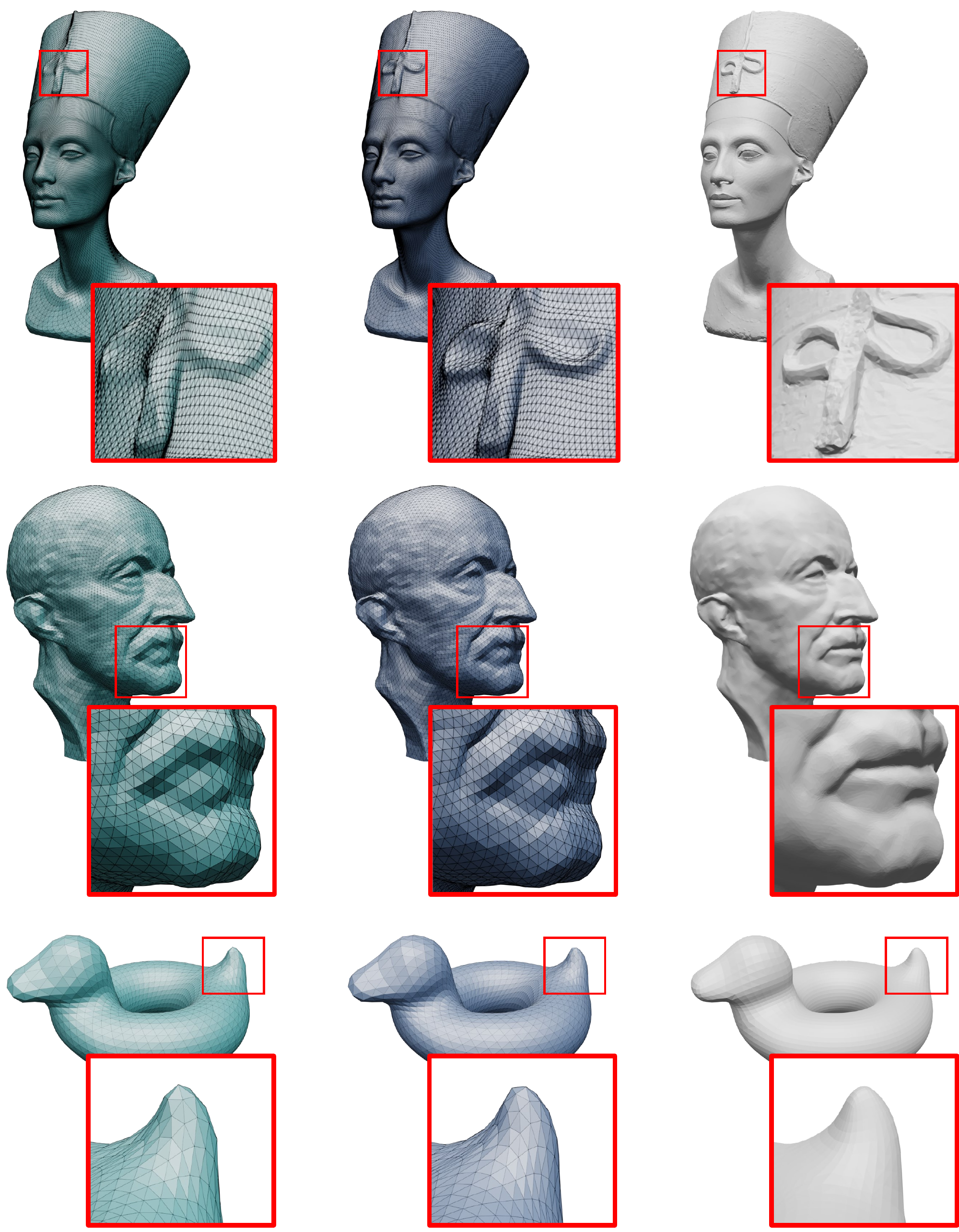}
\caption{Additional visual results for inverse rendering.}

    \label{fig:comparison-more}
\end{figure*}

\begin{figure}[htbp]
    \centering
    \includegraphics[width=0.48\textwidth]{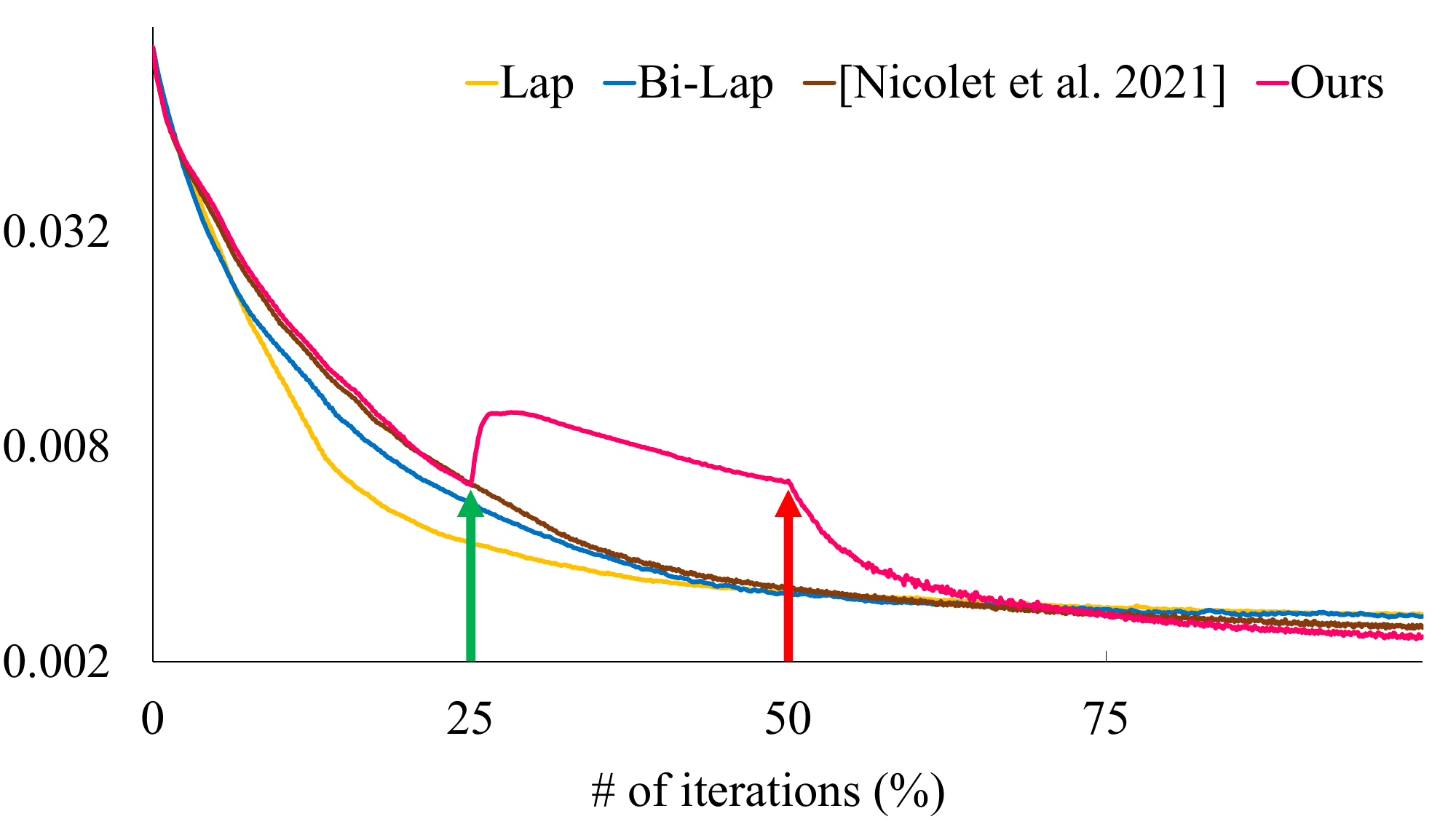}
    \caption{Progression of energy (MAE) over iterations in the inverse rendering experiment. Average energy is shown at each percentile. The Y axis is log-scaled using base $4$. In ours, introducing non-zero $w_k$ at 25\% increases the energy (green arrow), but by setting $w_u=0$ and $w_k=0$ after 50\% (red arrow), ours ends up with the lowest energy due to improved mesh density.}
    \label{fig:error-plot}
\end{figure}

We provide more results for inverse rendering in \Fig{comparison-more}.
We show the results of \cite{nicolet2021large} and ours using the same template mesh. \Fig{error-plot} further visualizes the progression of optimization over time. Reported photometric mean absolute error (MAE) values are averaged using the results from scenes of Suzanne, Cranium, Bob, Bunny, T-shirt, and Plank. The weight scheduling for the density adaptation term increases the error of our method at 25\% iteration mark, but due to better vertex density from the adaptation, our method results in the lowest error in the end.

\section{Additional results for non-rigid registration}
\label{sec:registration}

\begin{figure*}[htbp]
    \centering
    \begin{tabular}{cc}
    \rotatebox{90}{\ \ \ \ \ \ \ \ \ \ \ \ \ GT\ \ \ \ \ \ \ \ \ \ \ \ \ \ \ \ \ \ \ \ \ \ \ \ \ \ \ \ Ours \ \ \ \ \ \ \ \ \ \ \ \ \ \ \ \ \ \cite{qiu20213dcaricshop}}
    & 
    \includegraphics[width=0.95\textwidth]{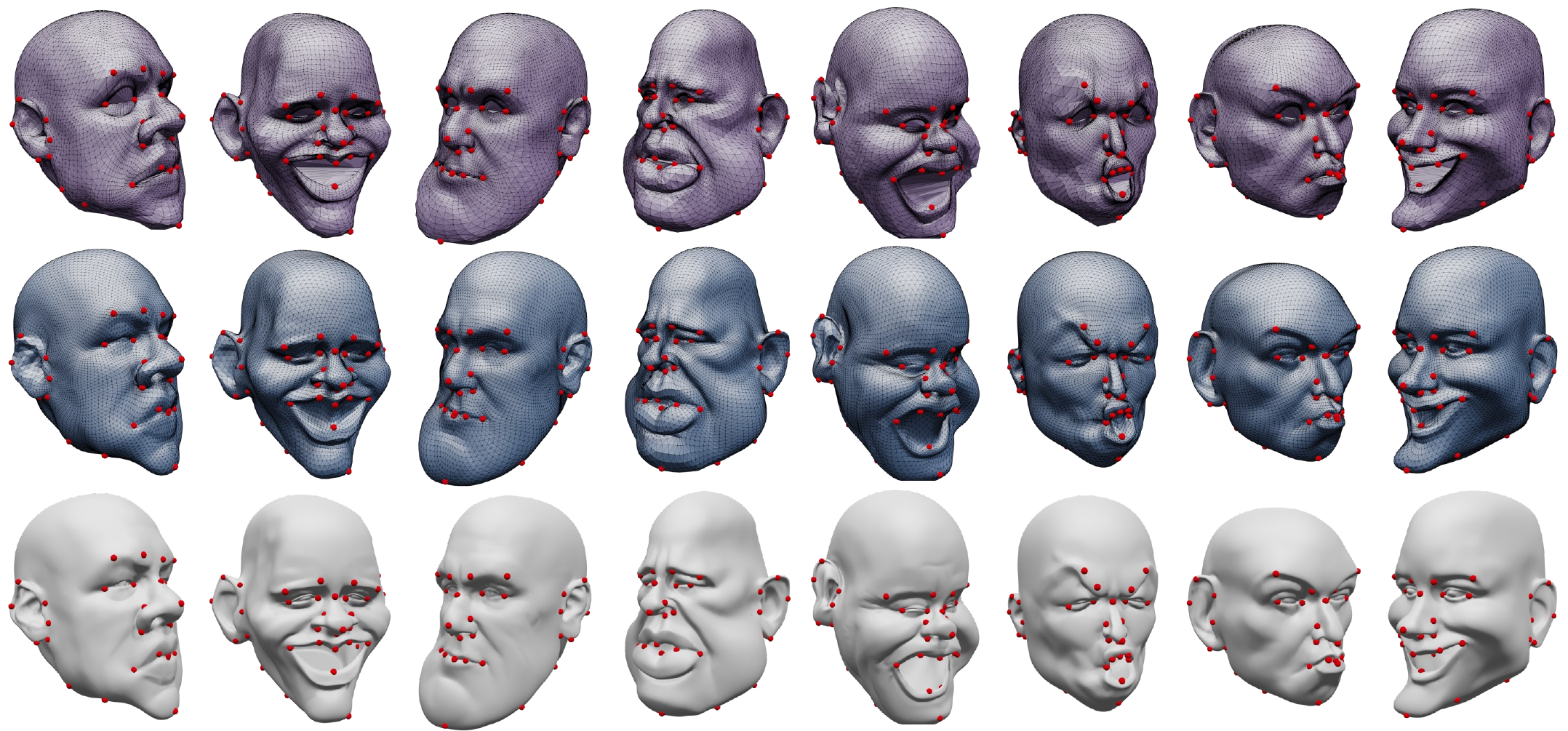}
    \end{tabular}
    
    \caption{Additional visual results of non-rigid registration.}

    \label{fig:non-rigid-results}
\end{figure*}

\Fig{non-rigid-results} presents more results from our non-rigid registration on the artist-sculpted 3D meshes in \cite{qiu20213dcaricshop}. Our non-rigid registration using the sphere template mesh with automatically annotated landmarks produces more accurate reconstructions.

\begin{figure}[htbp]
    \centering
    
    \begin{tabular}{c}
    \includegraphics[width=0.48\textwidth]{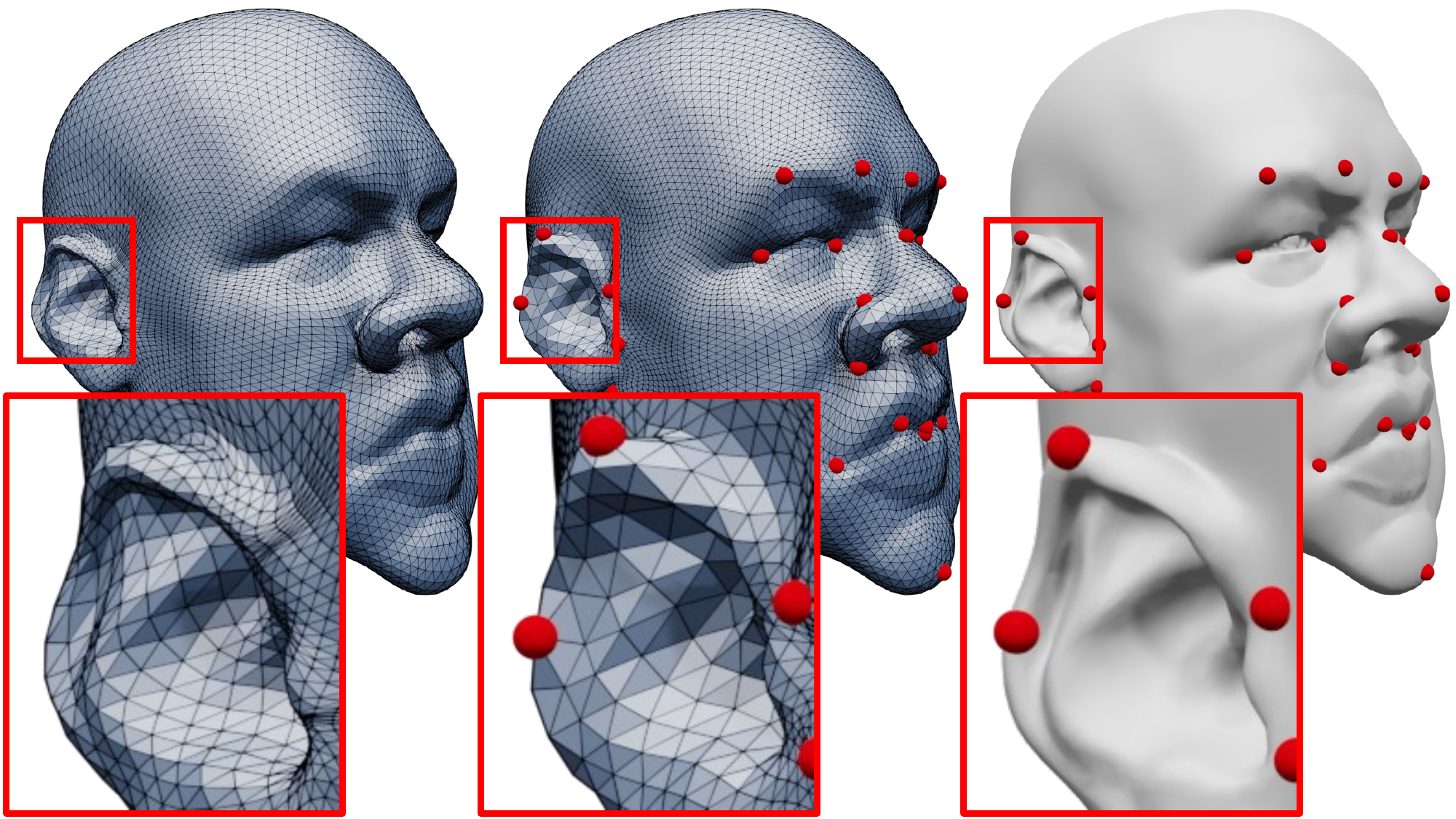} \\

    \begin{tabular}{ccc}
    
  {Ours w/o lmk \ \ \ \ \ \ \ \ \ \ \ \ \ \ Ours w/ lmk \ \ \ \ \ \ \ \ \ \ \ \ \ \ \ \ \ \ \ \ \ \ \ GT}\\
    \end{tabular}
    
    \end{tabular}
    \caption{Side-effect of using template landmarks. In some instances, using our template landmarks may hinder vertex relocation.}
    \label{fig:landmarks}
\end{figure}

Template landmarks may hinder vertex relocation needed for density adaptation, as shown in \Fig{landmarks}. Configuration of landmarks affects the resulting mesh density; vertices would not freely relocate around landmarks because the energy term for the landmarks in the optimization pins the 3D coordinates of landmark vertices on the template. So, compared to without-landmark fitting, fitting with landmarks may exhibit sparser density in some regions. This side effect can be avoided by fitting with a fewer number of landmarks or without landmarks.

\begin{figure*}[htbp]
    \centering
    \includegraphics[width=0.98\textwidth]{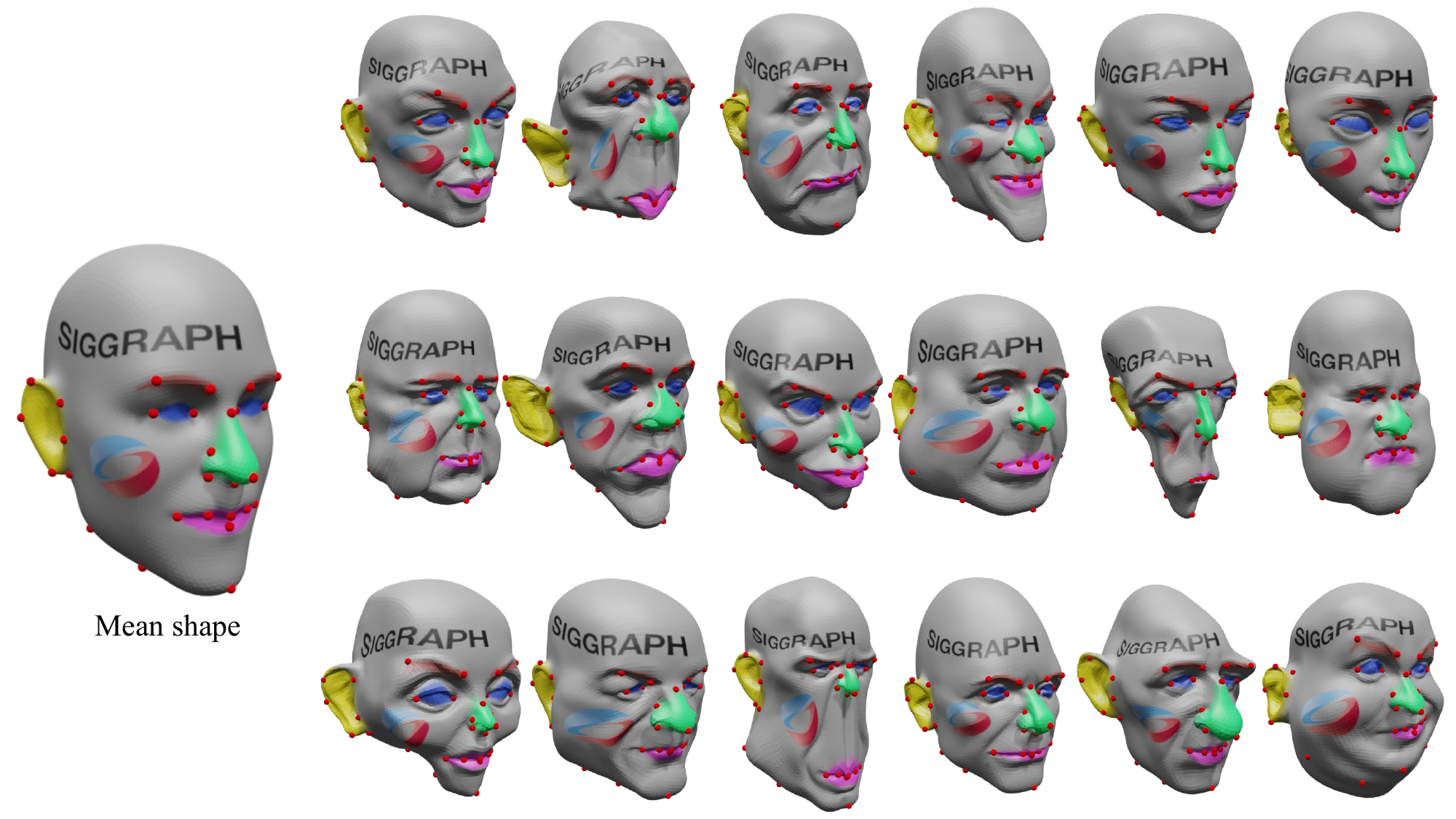}
    \caption{Texture transfer results using dense correspondences obtained from our non-rigid registration.}
    \label{fig:texture-transfer}
\end{figure*}

Using the dense correspondence obtained from our non-rigid registration, texture maps can be shared between the registration results, enabling texture transfer.
We show texture transfer results in \Fig{texture-transfer}.

\section{Limitations}
\label{sec:limitation}

\begin{figure}[htbp]
    \centering
    \includegraphics[width=0.48\textwidth]{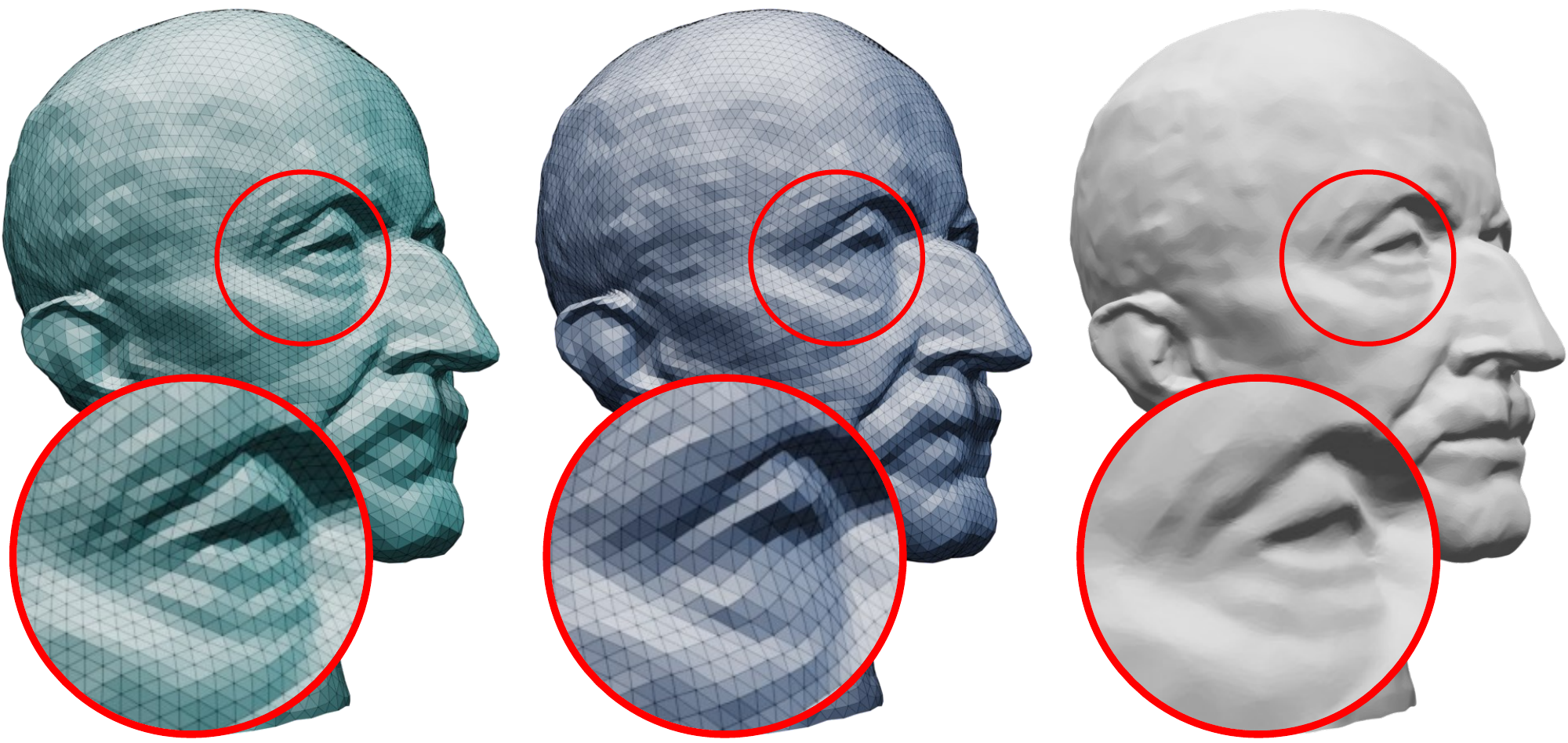}
    \caption{Limitations.
The vertex density for fine-scale details on a flat region may not increase. Large extruded regions are manifested early in the iterative optimization, so our density adaptation attracts vertices to such regions. In this example, the density around the extruded chin is increased, but the density increase is accompanied by decreases in the density of other regions, such as the eyes. Fine details manifest late in the optimization, so vertices may not be strongly attracted to them. Left: Without density adaptation. Middle: With density adaptation. Right: Ground truth.}
    \label{fig:limitation}
\end{figure}

\Fig{limitation} visualizes the limitation of our density adaptation based on Laplacian calculation from intermediate optimization results. While our proposed algorithm for calculating desired mesh density handles large structures well, fine details on flat region may not be benefited from the increase in density for better reconstruction.

\newpage
\bibliographystyle{ACM-Reference-Format}
\bibliography{main}

\clearpage

\clearpage

\clearpage

\newpage

